\documentclass[a4paper,11pt]{article}
\pdfoutput=1 

\usepackage{jheppub} 
\makeatletter
\DeclareRobustCommand*{\bfseries}{%
  \not@math@alphabet\bfseries\mathbf
  \fontseries\bfdefault\selectfont
  \boldmath
}
\makeatother

\usepackage{calc}
\usepackage{rotating}
\usepackage[english]{babel}
\usepackage[utf8]{inputenc}
\usepackage{graphicx}
\usepackage{subfig}
\usepackage{float}
\usepackage{amsmath}
\usepackage{amssymb}
\usepackage{amsthm}
\usepackage{latexsym}
\usepackage{dcolumn}
\usepackage{hyperref}
\def\gmtwo{(g-2)_\mu}

\def\gtilde{\tilde{g}}
\def\gvmu{g^V_{\mu}{}}
\def\gamu{g^A_{\mu}{}}

\def\ms{m_{\Phi}}
\def\mv{m_V}

\def\mfi{m_{\chi_i}}
\def\mmu{M_\mu}
\def\mks{M_{K^*}}
\def\mB{M_{B}}
\def\mBs{M_{B_s}}
\def\mK{M_{K}}

\restylefloat{figure}

\title{Flavour anomalies from a split dark sector}

\author[a]{Luc Darm\'e,}
\author[b]{Marco Fedele,}
\author[c]{Kamila Kowalska}
\author[c]{and Enrico Maria Sessolo}

\affiliation{$^a$ INFN, Laboratori Nazionali di Frascati,\\ C.P. 13, 100044 Frascati, Italy\\
$^b$ Dept.~de F\'{\i}sica Qu\`antica i Astrof\'{\i}sica, Institut de Ci\`encies del Cosmos (ICCUB),\\ Universitat de Barcelona, Mart\'i i Franqu\`es 1, E-08028 Barcelona, Spain\\
$^c$ National Centre for Nuclear Research,\\ ul.~Pasteura~7, 02-093 Warsaw, Poland}

\emailAdd{luc.darme@lnf.infn.it}
\emailAdd{marco.fedele@icc.ub.edu}
\emailAdd{kamila.kowalska@ncbj.gov.pl}
\emailAdd{enrico.sessolo@ncbj.gov.pl}

\abstract{We investigate solutions to the flavour anomalies in $B$ decays based on loop diagrams of a ``split''
dark sector characterised by the simultaneous presence of heavy particles at the TeV~scale and light particles around and below the $B$-meson mass scale.
We show that viable parameter space exists for solutions 
based on penguin diagrams with a vector mediator,
while minimal constructions relying on box diagrams are in strong tension with the constraints from 
the LHC, LEP, and the anomalous magnetic moment of the muon.
In particular, we highlight a regime where the mediator lies close to the $B$-meson mass, naturally realising a resonance structure and a $q^2$-dependent effective coupling. 
We perform a full fit to the relevant flavour observables and analyse the constraints from intensity frontier experiments. 
Besides new measurements of the anomalous magnetic moment of the muon, we find that decays 
of the $B$~meson, $B_s$-mixing, missing energy searches at Belle-II, and LHC searches for top/bottom 
partners can robustly test these scenarios in the near future.}

\begin{document}
\maketitle
\section{Introduction\label{sec:intro}}

Several flavour anomalies have been observed in the last few years in various $B$-meson decay measurements by different experimental collaborations. Some of the anomalous measurements involve semileptonic $b\to s$ transitions and include: (1) the suppression with respect to the Standard Model (SM) expectation of the ratios $R_{K}$ and $R_{K^{\ast}}$ -- the branching ratios of the $B$-meson decay into a $K$ or $K^*$ meson and muons, over the decay to the same kaon and electrons -- which were observed at LHCb~\cite{Aaij:2014ora,Aaij:2017vbb,Aaij:2019wad,Abdesselam:2019wac} and which imply the possible 
violation of lepton-flavour universality (LFUV);
(2) an enhancement in the angular observable $P_5'$, first measured by the LHCb~\cite{Aaij:2015oid} and Belle collaborations~\cite{Wehle:2016yoi}, and later observed also by ATLAS and CMS~\cite{Sirunyan:2017dhj,Aaboud:2018krd}; and (3) a suppression in the branching ratios for the decays $B_s\to\phi\mu^+\mu^-$\cite{Aaij:2015esa} and $B\to K^{(\ast)}\mu^+\mu^-$\cite{Aaij:2014pli,Aaij:2016flj}.

Global effective field theory analyses of the $b\to s$ data have pointed strongly towards New Physics~(NP) in the four-fermion operators $\mathcal{O}_9^{(\prime)\mu},\mathcal{O}_{10}^{(\prime)\mu}$~\cite{Altmannshofer:2014rta,Altmannshofer:2017fio,Capdevila:2017bsm,Altmannshofer:2017yso,DAmico:2017mtc,Ciuchini:2017mik,Alok:2017sui,
Hurth:2014vma,Hurth:2016fbr,Chobanova:2017ghn,Hurth:2017hxg,Arbey:2018ics,Alguero:2019ptt,Alok:2019ufo,Ciuchini:2019usw,Datta:2019zca,Aebischer:2019mlg,Kowalska:2019ley,Arbey:2019duh,Bhattacharya:2019dot}. Different combinations of the corresponding Wilson coefficients seem to be equally favoured, 
as long as $C_9^{\mu}$ remains significantly below zero. 
For instance, in a single operator scenario involving only $\mathcal{O}_9^{\mu}$, 
a solution consistent with the full set of $b\to s$ measurements 
requires approximately $C_9^{\mu}\in[-1.2,-0.6]$ at the $2\,\sigma$ level.

While the most natural assumption is that heavy
states above the scale of electroweak symmetry breaking (EWSB) 
are responsible for generating these operators, alternative possibilities have emerged in the 
literature~\cite{Datta:2017pfz,Sala:2017ihs,Alok:2017sui,Datta:2017ezo,Altmannshofer:2017bsz,Datta:2019bzu}, that these NP effects are in fact due to the presence of light degrees of freedom around or below the typical scale proved by the experiment.

The solutions based on a light dark sector can be divided roughly into two categories, depending on whether the masses involved lie above or below
the characteristic window for LFUV observables, which 
is commonly identified as ranging roughly from the dimuon threshold 
to $\sqrt{6}-\sqrt{8}\,\textrm{ GeV}$.
The first category involves 
a new light vector with mass $m_V \gtrsim 2 \,\textrm{GeV}$, 
coupled to the $b-s$ and the muon currents, 
interfering negatively with the SM amplitude~\cite{Sala:2017ihs,Alok:2017sui}.  
The light particle can contribute strongly to the physical observables thanks to the proximity of a resonance to the experimental bins~\cite{Sala:2017ihs} and NP effects can be parametrised in this case by Wilson coefficients with an explicit $q^2$ dependence. Interestingly, the corresponding resonance in the dimuon spectrum from $B$-meson decay could be hidden due to the sizeable hadronic uncertainty and the presence of the $J/\Psi$ resonance in the same region~\cite{Sala:2017ihs,Lyon:2014hpa}.

A second class of models, featuring instead the exchange of light particles below the dimuon threshold, have been considered in refs.~\cite{Datta:2017ezo,Alok:2017sui,Datta:2017pfz,Altmannshofer:2017bsz}. 
These scenarios are strongly constrained by a number of observations. On the one hand, a light scalar particle with effective couplings to the $s$ and $b$ quarks and leptons yields a positive contribution to the decay rate. One thus requires a sizeable coupling to electrons, a scenario that is in most cases~\cite{Datta:2017ezo} in tension with the measurement of $B \rightarrow K^{(\ast)} e^+ e^-$ processes in agreement with the SM~\cite{Aaij:2015dea}. On the other hand, a light vector boson with effective couplings to the $s$ and $b$ quark and muons can interfere negatively with the SM process, but is strongly constrained by the measurement of the anomalous magnetic moment $(g-2)_{\mu}$. This in turn implies a sizeable coupling to the $b$ and $s$ quarks, leading to strong bounds from $B_s$ mixing.
Finally, in ref.~\cite{Altmannshofer:2017bsz} it was pointed out that a resonance coupling to electrons can actually be used  to reproduce the low-$q^2$ bin of $R_{K^{\ast}}$. A vector boson very close -- but below -- the dimuon threshold, where $\mathcal{BR}(B\rightarrow K^{(*)} V)$ can be as small as $1\times 10^{-7}$, can explain the anomaly and escape the limits from ref.~\cite{Aaij:2015dea}.

A common trait of the constructions mentioned above is the presence of an effective off-diagonal coupling $g_{bs}$ to the quark current. Since the quarks carry SU(3)$_c$ charge, 
$g_{bs}$ must arise from particles with colour
integrated out in the UV theory, which must be heavy to avoid exclusion by the strong limits from the LHC. 
This also means that in realistic scenarios $g_{bs}$ will be suppressed by either loop effects, a small tree-level mixing angle between the SM quarks and heavy vector-like (VL) particles, or a combination of both.

In this paper we perform a detailed exploration of the first of these possibilities, i.e, the UV-complete model gives 
rise to the effective
$b-s$ coupling of the light particle via loop effects. 
We will show that, under these assumptions, in order to generate a $g_{bs}$ 
large enough to fit the flavour anomalies and, at the same time, 
maintain reasonable agreement with perturbativity of the couplings,
several phenomenological challenges must be faced.
In the specific of the cases mentioned above, 
we will show that solutions with $m_V \gtrsim 2\,\textrm{GeV}$
can be made viable with appropriate UV completions, 
even though they are subject to a combination of increasingly tightening bounds that include, in the UV,
LHC direct constraints on hadronic and leptonic new states and Drell-Yan dimuon constraints from the $Z$ lineshape, 
and in the IR, an appropriate width- and bin-dependent treatment of the bounds from $\mathcal{BR}(B\to K+\textrm{invisible})$ and $\mathcal{BR}(B\to K\mu\mu)$. 
Conversely, the second class of solutions, 
characterised by $m_V$ under the dimuon threshold,
is less appealing, as is strongly constrained by a combination of bounds from $\mathcal{BR}(B\to K+\textrm{invisible})$, 
neutrino trident production, intensity frontier limits on kinetic mixing, and $\mathcal{BR}(B_s\to \mu \mu)$.

More in general, we perform in this work
a detailed Monte Carlo scan of the broad $m_V$ range and the loop-induced couplings. 
We identify and characterise the specific properties of the viable models 
and provide an indication of possible strategies for a timely detection of the associated NP states.

The paper is organised as follows. In Sec.~\ref{sec:oneloop} we recall expressions for the loop-generated Wilson coefficients from box and penguin diagrams. We provide the quantum numbers of the particles entering the loops and estimate the characteristic size of the couplings required to fit the flavor anomalies. In Sec.~\ref{sec:B-anom} we enlist the complete set of constraints we apply to the models. Section~\ref{sec:results} is dedicated to the main results, with a description of the fitting procedure and discussion of the allowed parameter space. We present our conclusions in Sec.~\ref{sec:concl}. Appendix~\ref{appen} is dedicated to the detailed treatment of the recasting procedure for $B \to K+\textrm{inv.}$ limits. 

\section{Effective one-loop Wilson coefficients from split dark sector models\label{sec:oneloop}}

Our goal is to investigate to what extent the $b-s$ flavour anomalies can be explained by generic 
loop effects involving \textit{light} new particles (in association with a heavy sector above the EWSB scale).
Possible constructions for the Wilson coefficients of the effective Hamiltonian descending from loops of TeV-scale new 
particles have been investigated, e.g., in refs.~\cite{Gripaios:2015gra,Arnan:2016cpy,Cline:2017qqu,Crivellin:2018yvo,Datta:2018xty,Barman:2018jhz,Marzo:2019ldg,Arnan:2019uhr}. 
They generally involve either box diagrams of 
scalar and fermionic states like in Fig.~\ref{fig:loopdiag}(a), or penguin diagrams like the ones represented in Fig.~\ref{fig:loopdiag}(b).  

\subsection{Box diagrams}

\begin{figure}[t]
\centering
\subfloat[]{%
\includegraphics[width=0.47\textwidth]{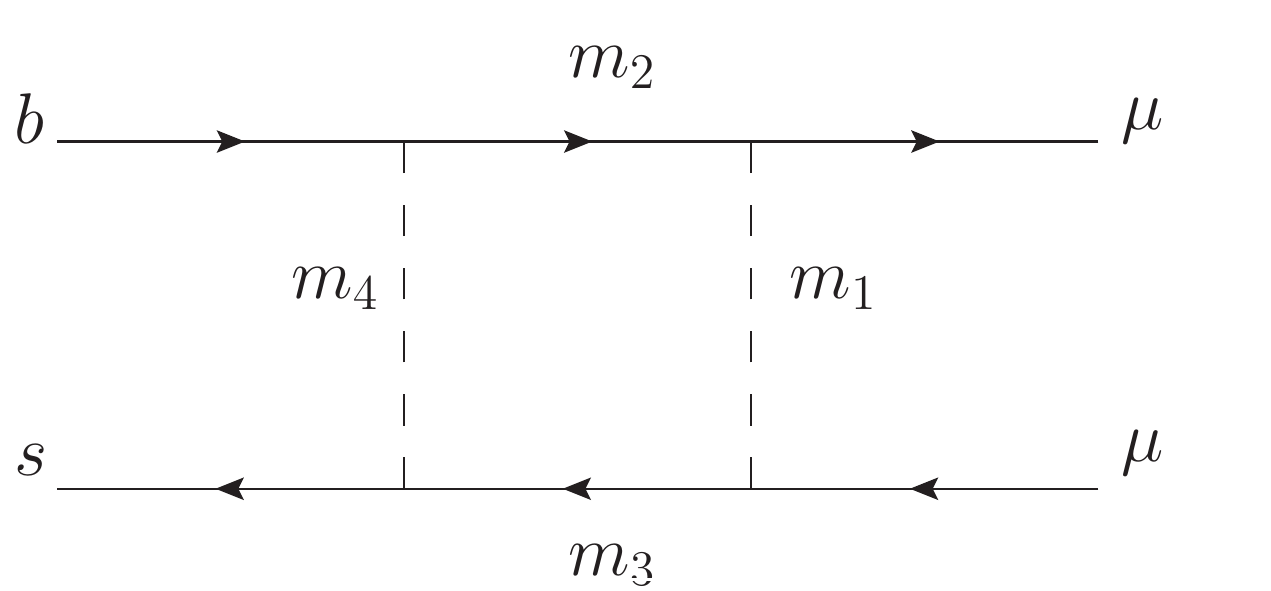}
}%
\hspace{0.02\textwidth}
\subfloat[]{%
\includegraphics[width=0.47\textwidth]{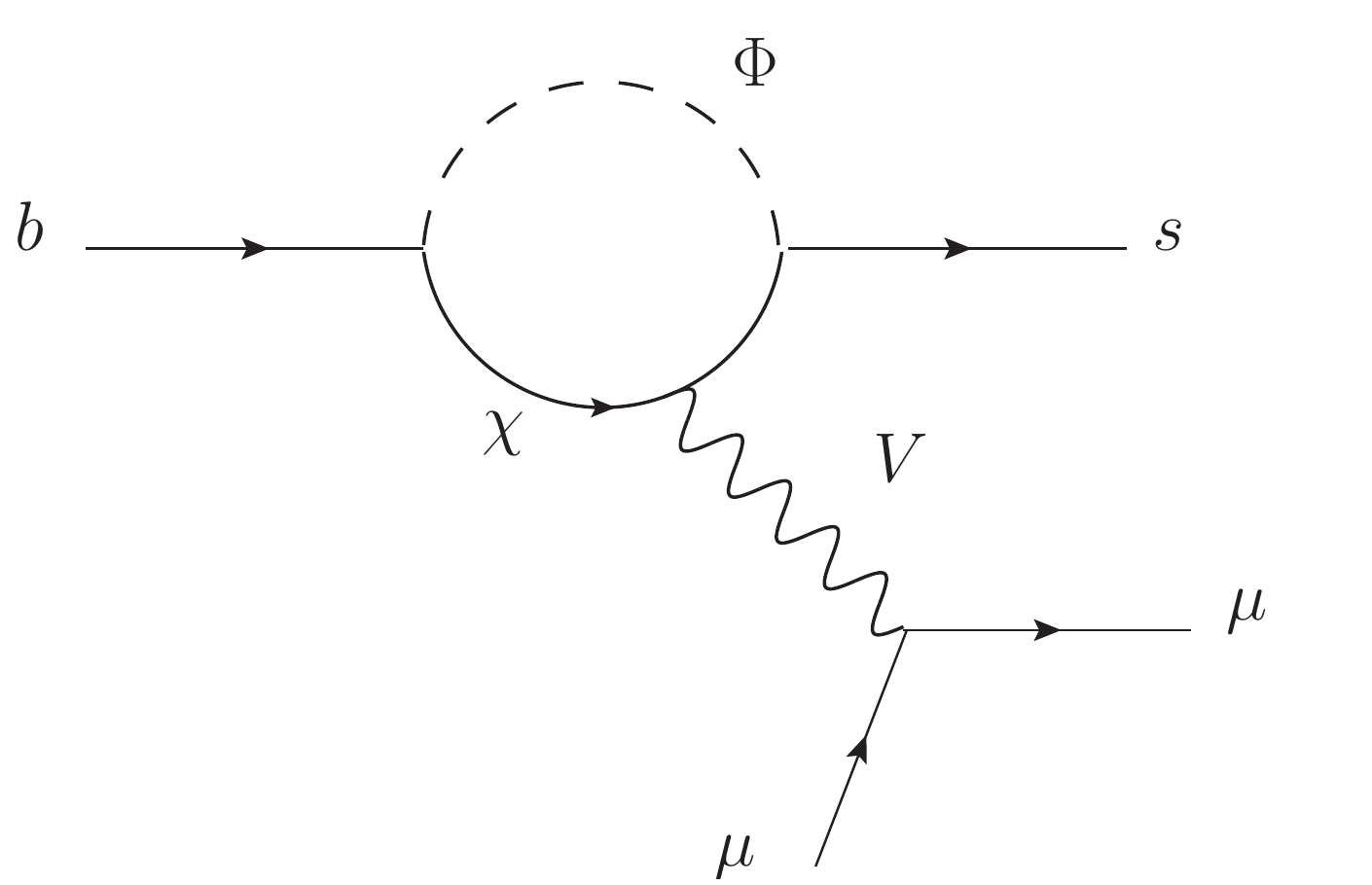}
}%
\caption{(a) Example box loop mediating the $B$-meson decay.
(b) Example penguin loop mediating the $B$-meson decay.}
\label{fig:loopdiag}
\end{figure}

If direct Yukawa couplings between the quarks $b,s$, the muons, and a NP sector 
composed of fermions $\psi_i$ and scalars $\phi_j$ are allowed by the gauge quantum numbers, 
the most generic Lagrangian is given by
\begin{multline}\label{boxlagr}
    \mathcal{L}\supset \bar{\psi}_i\left(Y_{L,ij}^{(b)} P_L b+Y_{L,ij}^{(s)} P_L s+Y_{L,ij}^{(\mu)} P_L \mu\right)\phi_j\\
    +\bar{\psi}_k\left(Y_{R,kl}^{(b)} P_R b+Y_{R,kl}^{(s)} P_R s+Y_{R,kl}^{(\mu)} P_R \mu\right)\phi_l+\textrm{H.c.}\,,
\end{multline}
where a sum over repeated indices is implied.
It was shown, e.g., in refs.~\cite{Gripaios:2015gra,Arnan:2016cpy,Arnan:2019uhr,Kawamura:2017ecz} that 
box diagrams like the one depicted in Fig.~\ref{fig:loopdiag}(a)
can then contribute to the $B\to K^{(\ast)}\mu^+\mu^-$ decay and
one can construct out of Eq.~(\ref{boxlagr}) Wilson coefficients $C_9^{(\prime)\mu}, C_{10}^{(\prime)\mu}$ 
of the right order of magnitude to fit the anomalies. On the other hand, these box-like constructions
do not generally involve  
light particles or very split spectra, as they would incur strong tension with several existing bounds.

Since at least one of the fermion or scalar fields in the box must necessarily carry the colour charge, 
the bounds from LHC searches for colour production with $b$-tagged jets will contribute
to push this state above the $1-1.2\,\textrm{TeV}$ scale~\cite{Aaboud:2017wqg,Aaboud:2017ayj,Sirunyan:2019glc,Sirunyan:2019ctn}.
At the same time there exist strong bounds from the measurement of $B_s$ mixing~\cite{DiLuzio:2017fdq}
that either limit the available size of the 
Yukawa couplings or push one of the new states coupled to $b,s$ to a very large scale. The strength of the $B_s$-mixing bound is
very model-dependent and, if the particle content of the model at hand is large enough, 
cancellations between different diagrams can be engineered to evade the limit. 
Nevertheless, we can easily obtain an estimate of the $B_s$-mixing bound in minimal cases, when the particle content is just about 
right to fit the flavour anomalies via \textit{a} box diagram similar to Fig.~\ref{fig:loopdiag}(a). Assuming that 
the colour charge is carried by particle~4 in the figure\footnote{The opposite choice would imply that particles 1, 2, and 3 carry colour, which in turn would lead to most states in the loop being either at the TeV scale or, in fact, SM quarks. While the latest possibility is interesting, it would trigger very strong tree-level flavour-violating bounds whose effect are likely to dominate the loop-induced signatures we consider here.} we get, following, e.g., ref.~\cite{Arnan:2016cpy},  
$|Y_{L,24}^{(b)}Y_{L,34}^{(s)\ast}|\lesssim 0.09$ at the $2\sigma$ level. 

If particle~4 carries colour, the other states can in principle be lighter: $m_1\approx m_2\approx m_3\lesssim m_4\approx \mathcal{O}(\textrm{TeV})$. The Wilson coefficient $C_9^{\mu}$ is then approximately calculated as (see, e.g., refs.~\cite{Arnan:2016cpy,Arnan:2019uhr})
\begin{equation}
    C_9^{\mu}\approx -\frac{Y_{L}^{(s)\ast}Y_{L}^{(b)}|Y_{L,R}^{(\mu)}|^2}{(m_4/\textrm{TeV})^2}F(x,y,z)\,,\label{C9box}
\end{equation}
where we have suppressed the subscript indices in the Yukawa couplings to lighten the notation, we define
$x=(m_1/m_4)^2$, $y=(m_2/m_4)^2$, $z=(m_3/m_4)^2$, and $F$ is a loop function,
\begin{equation}
F(x,y,z)=\frac{x^2\ln x}{(x-1)(x-y)(x-z)}
+\frac{y^2\ln y}{(y-1)(y-x)(y-z)}
+\frac{z^2\ln z}{(z-1)(z-x)(z-y)}\,,\label{boxloop}
\end{equation}
which equals approximately~1 when $m_1\approx m_2\approx m_3\approx \mathcal{O}(100\textrm{ GeV})$.
As was shown in, e.g., refs.~\cite{Gripaios:2015gra,Arnan:2016cpy,Arnan:2019uhr,Kawamura:2017ecz}, it follows from Eq.~(\ref{C9box}) that
$C_9^{\mu}\lesssim -0.6 $ requires Yukawa couplings $|Y_{L,R}^{(\mu)}|\gtrsim 3$ for $m_1\approx m_2\approx m_3\approx \mathcal{O}(100\textrm{s GeV})$, 
a condition that does not guarantee the validity of perturbation theory up to scales much larger than EWSB.

One might wonder at this point if an eventual light sector in the theory can provide a more natural fit to the flavour anomalies, perhaps requiring smaller Yukawa couplings to the muons. 
In fact, one infers from Eq.~(\ref{boxloop}) that $F$ can receive a logarithmic enhancement of a few units if 
$x$, $y$, and $z$ are all at the same time significantly smaller than~1. But
at least one among $m_1$, $m_2$, $m_3$ cannot be much smaller than $m_4$.
There are two reasons for this. First and foremost, multi-lepton searches at the LHC via Drell-Yan production constrain the particles 
carrying SU(2)$_L\times$U(1)$_Y$ quantum numbers 
to masses above the $400-600\,\textrm{GeV}$ 
range~\cite{Kowalska:2017iqv,Sirunyan:2018lul,Aad:2019vnb}. On the other hand, if one were to roughly extrapolate 
a similar reasoning to  $m_1$, $m_2$, $m_3$ smaller than the mass of the muon, so that the corresponding particles may possibly result invisible at the LHC, these
would still necessarily contribute at one loop to the anomalous magnetic moment of the muon. 
The measured $2 \sigma$ upper bound, $\delta(g-2)_{\mu}\lesssim 4\times 10^{-9}$~\cite{Bennett:2006fi,Davier:2016iru,Jegerlehner:2017lbd} 
implies that the muon Yukawa coupling cannot be larger than 
\begin{equation}\label{eq:yukg2}
    |Y_{ij,L(R)}^{(\mu)}| \lesssim 10^{-2}\left( \frac{m_{1,2,3}}{\textrm{GeV}}\right),
\end{equation}
quite independently of the specifics of the model at hand.

A rough comparison of the typical size of Eq.~(\ref{C9box}) and Eq.~(\ref{eq:yukg2}) shows that the latter is 
too small to fit the flavour anomalies unless $m_{1,2,3}$ lie in the few hundreds of~GeV or above.
We conclude that there is arguably 
no common parameter space for $b\to s$ anomalies and $(g-2)_\mu$ with perturbative Yukawa couplings and a minimal, light NP sector, if the contributions to $B$ decays stem from this class of box diagrams. 

\subsection{Penguin diagrams}
\label{sec:penguin}
It is more promising to look at another type of loop-induced coupling giving rise to the effective operators $\mathcal{O}_9^{(\prime)},\mathcal{O}_{10}^{(\prime)}$: 
the penguin-diagram generated interaction.
The penguin is constructed out of a loop with new fermion and scalar fields exchanging a vector boson with the leptons. If the vector boson is one of the SM gauge bosons, the contribution to $C_9^{(\prime)}$, $C_{10}^{(\prime)}$ is flavour-universal, and cannot be used to explain the LFUV anomalies.
We will therefore be interested in models realising the penguin diagram topology presented in Fig.~\ref{fig:loopdiag}(b), in which a new light gauge boson $V$ couples to the muons. 

The particle that carries the colour charge in the loop might be a scalar multiplet $\Phi$ or a heavy VL fermion $\Psi$. In the former case we close the loop with a light (Dirac) fermion $\chi$, whereas in the latter with a light scalar $\pi$. 
In order to avoid 
charging the $b$ and $s$ quarks under the dark gauge group U(1)$_D$ (with gauge coupling $g_D$), 
we assume that the dark charge is confined within the loop, i.e., $Q_{\chi(\pi)}=-Q_{\Phi(\Psi)}$. We thus avoid the strong bounds on
$V$ from multi-lepton searches at the LHC. 

Without much loss of generality we will focus henceforth 
on the case where the heavy coloured particle is $\Phi$. 
The case with $\Psi$ does not present very significant differences,
barring an order-one suppression of the amplitude  that comes from swapping 
the role of the light and heavy mass in the loop functions.   
We thus introduce a scalar doublet $\Phi=(\phi_u,\phi_d)^T$ and a few light fermion singlets $\chi_i$, 
whose  multiplicity will be specified case by case.  Explicitly, their
SU(3)$_c\times$SU(2)$_L\times$U(1)$_Y\times$U(1)$_D$ quantum numbers read
\begin{align*}
\Phi=(\mathbf{3},\mathbf{2},1/6,Q_{\Phi})\quad \quad \chi_i=(\mathbf{1},\mathbf{1},0,-Q_{\Phi}) \,.
\end{align*}
Note that with the above charge assignment the mass matrix of the dark fermions is not a priori diagonal, 
unless the off-diagonal entries are forbidden by a flavour symmetry, or suppressed by some other mechanism. 
We will assume for simplicity that this is always the case, without entering in the specifics of such constructions.   

We confine ourselves to the treatment of left-handed $b-s$ currents, in agreement with the findings of the global fits in the literature.
Below the EWSB scale the Lagrangian of the hadronic NP sector reads
\begin{equation}
\mathcal{L} \supset  y_d^{ij}\, \phi_d^{\ast}\, \bar{\chi}_i P_L d_j  +   y_u^{ij}\, \phi_u^{\ast}\,\bar{\chi}_i P_L u_j+\textrm{H.c.}\,,
\label{eq:L_B}
\end{equation}
where a sum over repeated indices is implied, and the Yukawa couplings are related by $y_u^{ij}=y_d^{ik}(V_{\textrm{CKM}}^{\dag})^{kj}$. We further confine ourselves to the basis where the only nonzero Yukawa 
couplings of the down-like type are those of the second and third generation: $y_d^{ik=1,2,3}=(0,y^i_s,y^i_b)^T$.

We finally introduce an effective 
interaction of the new gauge boson to the muons:
\begin{equation}
\mathcal{L} \supset \left( g^V_\mu \bar{\mu} \gamma_\nu \mu + g^A_\mu \bar{\mu} \gamma_\nu \gamma_5 \mu  \right) V^\nu.
\label{eq:L_V}
\end{equation}

The relative size of the $\gvmu$ and $\gamu$ couplings is governed by the UV model building and
so is the eventual size, if nonzero at all, of the coupling of the new gauge boson to neutrinos. 
This point is of particular relevance when it comes to the constraints on the model from neutrino trident production~\cite{Altmannshofer:2014pba} at
CCFR~\cite{Mishra:1991bv} and CHARM-II~\cite{Geiregat:1990gz}, 
to which we come back in Secs.~\ref{sec:B-anom} and \ref{sec:results}. 
We shall see that, while it is desirable to embed the effective model in a framework 
with forbidden or strongly 
suppressed couplings to the neutrinos,\footnote{For example, one can construct a Type-I 2-Higgs doublet model
where the additional Higgs doublet and extra VL heavy leptons all carry U(1)$_D$ charge:
$\Theta:(\mathbf{1},\mathbf{2},1/2,Q_{\Theta})$, $E:(\mathbf{1},\mathbf{1},1,Q_{\Theta})$. Yukawa couplings with the SM left-handed 
muon doublet $L_{\mu}$, of the type $\lambda\Theta^{\dag}L_{\mu}E$, generate a left-chiral coupling~$g_{\mu}^L$ of~$V$ to the muons once 
U(1)$_D$ is broken. The coupling to neutrinos is absent. Typically one gets $g_{\mu}^L\approx -g_D Q_{\Theta}\lambda v_{\Theta}^2/2 
m_E^2$. More than one family of VL singlet fermions, and an additional complex scalar singlet charged under U(1)$_D$ can be 
introduced to generate the right-chiral couplings and to push the mass of the scalars with electric charge above the current bound from 
LEP and LHC searches~\cite{Ko:2013zsa}.\label{foot:uv}} we are able to find 
in our numerical analysis some viable parameter space lying below the bound from CCFR and CHARM-II.  

We calculate the penguin diagram contributions to the $B_s\to K^{(\ast)}\mu^+\mu^-$ decay. Following standard techniques one obtains an amplitude that closely resembles the SM photon penguin case, with a slight modification due to the Breit-Wigner distribution of the massive gauge boson $V$.
Estimating the loop diagram is equivalent to integrating out the heavy colour-charged field, 
which generates the dimensionful Wilson coefficient $\tilde{g}$ of the operator 
\begin{equation}
\label{eq:O6op}
\mathcal{O}_6=\left(\bar{s}\gamma_{\rho}P_L b\right)\partial_{\sigma}V^{\rho\sigma}\,, 
\end{equation} 
where $V^{\rho\sigma}=\partial^{\rho} V^{\sigma}-\partial^{\sigma} V^{\rho}$~\cite{Altmannshofer:2017bsz,Datta:2017ezo}. 
We find 
\begin{equation}\label{eq:gtilde}
       \tilde{g} = -\frac{g_D\,Q_{\Phi}}{16 \pi^2 \ms^2 } \sum_i y_s^{i\ast} y_b^i\, \mathcal{F}(x_i)\, ,
\end{equation}
where we have defined $x_i = \mfi^2 / \ms^2$ and the loop function reads
\begin{equation}
\mathcal{F}(x)=-\frac{1}{2}-\frac{1}{3}\ln x\, .
\end{equation}

The contribution of $\mathcal{O}_6$ to the $B \to K$ processes mediated by the exchange of a vector boson can be 
related to ``effective'' Wilson coefficients $C_{9,10}^\ell$ as
\begin{equation}
\label{eq:Wilson}
C_{9(10)}^{\mu} (q^2) = -\frac{4\pi \mathcal{N}}{\alpha_{\textrm{EM}}} \tilde{g}\,g_{\mu}^{V(A)} \frac{q^2}{q^2-\mv^2+i m_V \Gamma_V} \,,
\end{equation}
where $m_V$ is the mass of the gauge boson, 
we define
\begin{align}
 \mathcal{N}^{-1} = \frac{4 G_F}{\sqrt{2}} V_{tb} V_{ts}^{\ast} \,,
\end{align}
and $\Gamma_V$ is the total width of the gauge boson, which reads, when all light decay channels are kinematically open,
\begin{equation}\label{eq:gammav}
\Gamma_V=m_V \left( \gamma_V^D + \gamma_V^\mu \right) \,,
\end{equation}
with the dark fermion and SM muon contributions respectively given by
\begin{align}
\gamma_V^D=&\frac{1}{12\pi}\left[ \sum_i \sqrt{1 - \frac{4 \mfi^2}{m_V^2}} \left( \frac{2\mfi^2}{m_V^2} + 1\right) g_D^2 Q_{\chi}^2 \right]\,, \label{eq:gammavD}\\
\gamma_V^\mu=&\frac{1}{12\pi}\left[ \sqrt{1 - \frac{4 M_\mu^2}{m_V^2}} \left(\frac{2 M_\mu^2}{m_V^2} + 1\right) \left(\gvmu\right)^2 
+  \left(1 - \frac{4 M_\mu^2}{m_V^2} \right)^{3/2} \left(\gamu\right)^2 \right] \label{eq:gammavmu}\,.
\end{align}
Note that the Wilson coefficient $\tilde{g}$, similarly to $C_{9(10)}(q^2)$,
does not run at the leading order in QCD, since its colour part is simply a vector current~\cite{Gracey:2000am}.

\begin{figure}[t]
	\centering
		\includegraphics[width=0.75\textwidth]{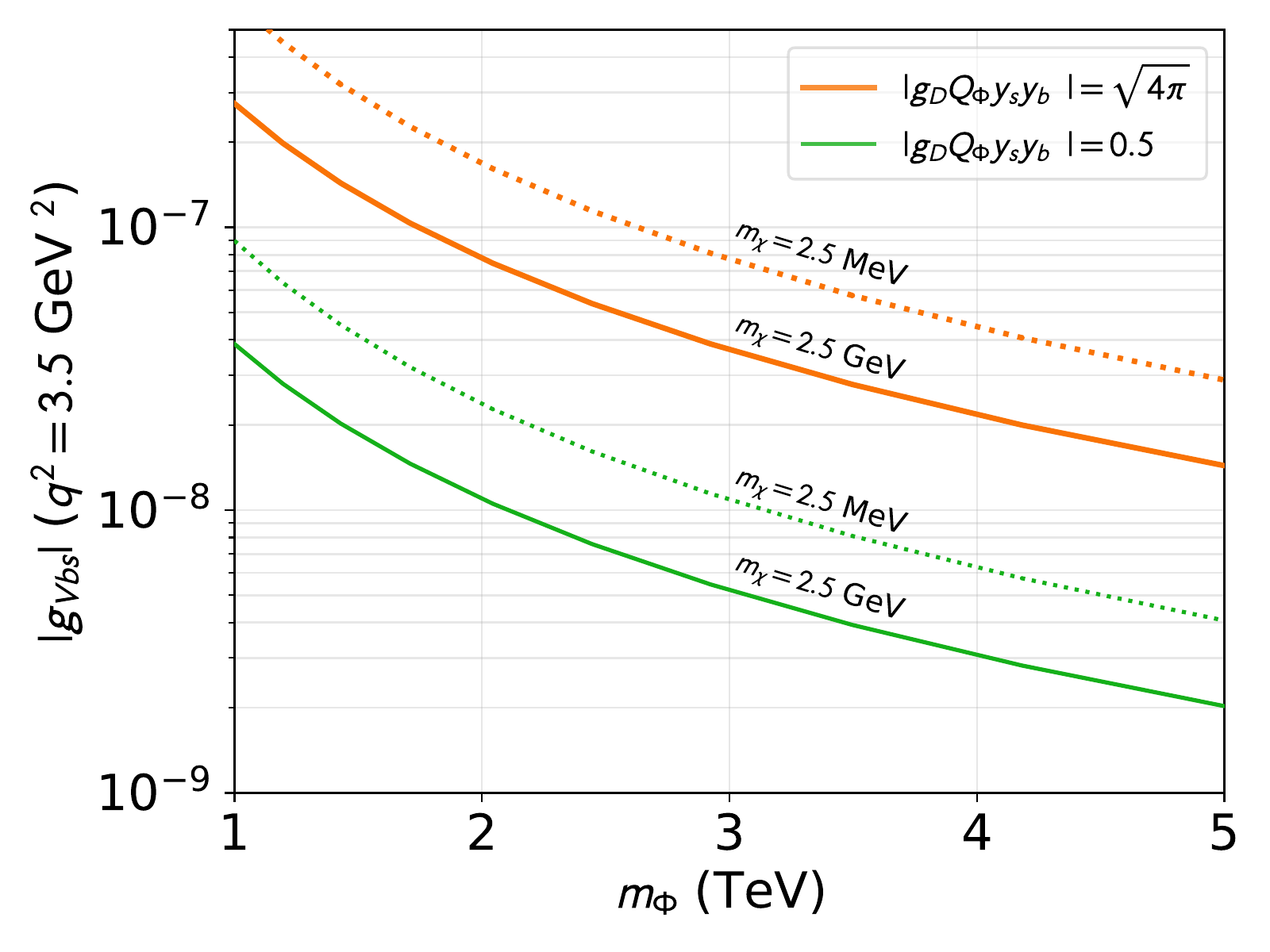}
	\caption{The typical size of $g_{Vbs}$ as a function of the heavy mass $m_{\Phi}$ for different choices of the effective coupling $|g_D Q_{\Phi} y_s^{i\ast} y_b^{i}|$ given here for one light dark fermion, $\chi_1$. Solid (dashed) lines correspond to $m_{\chi_1}=2.5\,\textrm{GeV(MeV)}$. Indicative momentum transfer is set at $q^2=3.5\,\textrm{GeV}^2$. }
			\label{fig:gVbs}
\end{figure}

To facilitate a comparison with the existing literature, we further define the dimensionless coupling 
of $V$ to the $b-s$ current, $g_{Vbs}$, as 
\begin{equation}\label{eq:gvbs}
        g_{Vbs} \equiv -\tilde{g}\,q^2\,.
\end{equation}
The typical size of $g_{Vbs}$ is shown in Fig.~\ref{fig:gVbs} for representative choices of the input parameters. 
For illustration purposes, we have set $q^2$
to the indicative scale of $3.5\,\textrm{GeV}^2$. Note, however, that in the phenomenological analysis of 
Sec.~\ref{sec:results} we integrate Eq.~(\ref{eq:Wilson}) over all the relevant invariant-mass bins.

\bigskip 

Equation~(\ref{eq:Wilson}) shows that the penguin-generated Wilson coefficients depend
on $q^2$ and the size of the width of the new gauge boson. They enter nontrivially in the calculation of the flavour observables and 
relative constraints when the $V$ mass lies in the vicinity of the experimental bins. The corresponding expressions should 
thus be compared directly to the experimental data, as we do in the numerical 
scan presented in Sec.~\ref{sec:results}.    
We can nonetheless provide a rough estimate of the typical size required for the dimensionful 
coupling $\tilde{g}$ in the limiting cases 
when $m_V$ 
becomes much larger or much smaller than the experimental energy (for the LFUV observables 
we indicatively consider this to be $q^2 \approx 0.04-8\,\textrm{GeV}^2$). If $m_V\gg 10\,\textrm{GeV}$ 
one obtains $C_9^{\mu}\lesssim -0.6$ roughly for
\begin{equation}\label{him_est}
|\tilde{g}\,g_{\mu}^V|\gtrsim 1 \times 10^{-10}\, \frac{m_V^2}{\textrm{GeV}^4}\,,  
\end{equation}
where we have set $q^2$ at the mean momentum transfer for the $1-6\,\textrm{GeV}^2$ bin 
of the $R_{K^{(\ast)}}$ observables.

The two couplings $\tilde{g}$ and $g_{\mu}^V$ 
are independently constrained by two powerful probes. 
On the one hand, the measured $2\sigma$ 
bound on $R_{BB}$ from $B_s$ mixing~\cite{DiLuzio:2017fdq} 
has a direct impact on $\tilde{g}$
when the vector $V$ is exchanged in the 
$s$-channel:\footnote{Note, however, that $g_{Vbs}$ depends 
directly on the product $y_s^{i\ast} y_b^i$, 
which produces box diagrams that are also contributing to $B_s$ mixing, 
and might result on a stronger bound than the one derived by $s$-channel 
exchange of $V$.
We will come back to this point in Sec.~\ref{sec:B-anom}.} 
\begin{equation}\label{gsbbound}
|\tilde{g}|\lesssim 7\times 10^{-8}\,\frac{m_V}{\textrm{GeV}^3}\,,
\end{equation}
where the characteristic experimental scale coincides with the $B_s$ mass,
$q^2=M_{B_s}^2$.

On the other hand, the already mentioned $2\sigma$ upper bound 
on the anomalous magnetic moment of the muon constrains $g_{\mu}^V$ to  
\begin{equation}\label{gmubound}
|g_{\mu}^V|\lesssim 7\times 10^{-3}\,\frac{m_V}{\textrm{GeV}}\,.
\end{equation}
The current bounds thus result in a very narrow window of availability for explaining the flavour anomalies with penguin diagrams and $m_V\gg 10\,\textrm{GeV}$. 

At the opposite side of the spectrum, $m_V< 200\,\textrm{MeV}$,
the new gauge boson is much lighter than the experimental energy scale. Equation~(\ref{eq:Wilson}) shows that the Wilson coefficients 
become independent of $m_V$ and of $q^2$. $C_9^{\mu}\lesssim -0.6$ requires approximately 
\begin{equation}\label{bou_prod_light}
|\tilde{g}\,g_{\mu}^V|\gtrsim 5 \times 10^{-10}\,\textrm{GeV}^{-2}\,.    
\end{equation}
The most severe constraint in this mass range comes again from the $2\sigma$ upper bound on 
the anomalous magnetic moment of the muon, which requires
\begin{equation}\label{gmublight}
|g_{\mu}^V|\lesssim 0.6-1\times 10^{-3}\,,
\end{equation}
where the upper value refers essentially only to the region $m_V\gtrsim 100\,\textrm{MeV}$, 
and the lower value to all other masses below that. 
In light of Eqs.~(\ref{bou_prod_light}) and (\ref{gmublight}) one needs
$|\tilde{g}|\gtrsim 10^{-6}\,\textrm{GeV}^{-2}$ 
to fit the flavour anomalies. As one can see in
Fig.~\ref{fig:gVbs}, effective couplings of this size are not easy to obtain in the penguin setup. 

It is well known (cf., e.g., ref.~\cite{Sala:2017ihs}) that one can increase the size 
of $\gvmu$ while respecting the $2\sigma$ upper bound on $\gmtwo$ by introducing 
the axial-vector coupling $\gamu$ and thus creating a negative contribution to $\gmtwo$ that has to be fine-tuned. This, however, also induces an extremely strong contribution to the $B_s \to \mu \mu$ decay rate, and will prove ultimately impracticable, as will appear clear in the next sections. 

Altogether, this discussion leads us to conclude
that the most natural solutions are likely to be situated inside the
window $\mv \approx 0.2-10\,\textrm{GeV}$. 
The remainder of this paper is thus dedicated mostly to this mass range. 
Note that when $m_V$ approaches $M_{B}$, additional resonant enhancement can be obtained to open up the parameter space, 
albeit at the cost of additional limits from $B \to K$ processes, which we will describe in detail in the next section.

\section{Constraints on the model\label{sec:B-anom}}

\subsection{Flavour constraints\label{sec:flav}}

\paragraph{$\boldsymbol{B_s}$\textbf{-mixing}}
Strong limits on the Yukawa couplings of Eq.~(\ref{eq:L_B}) arise from 
box-diagram contributions to $B_s$-mixing. We recall that for exclusively left-handed couplings the only relevant operator is $\mathcal{O}_1$~\cite{Arnan:2016cpy,Arnan:2019uhr}. 
The corresponding dimensionful Wilson coefficient is defined as
  \begin{equation}
C_1 =  \frac{1}{128 \pi^2 \ms^2}  \sum_{ij} y_s^{i\ast} y_b^i  y_s^{j\ast} y_b^j  F (x_i,x_j)\,,   
\end{equation}
where the sum runs over all possible dark fermions $\chi_{i,j}$, out of which we can construct a box diagram with $\Phi$. We define 
 $x_{i(j)}=m_{\chi_{i(j)}}^2/m_{\Phi}^2$ and the loop function reads 
 \begin{equation}
F(x,y) = \frac {1}{(1-x)(1-y)}  +\frac{x^2 \log x}{(1-x)^2(x-y)}
 +\frac{y^2 \log y}{(1-y)^2 (y-x)}\,.
\end{equation}
 
In the small $x,y$ limit, the loop function can be approximated as $F(x,y)  \simeq 1 $,
so that  $C_1 \propto 1/\ms^2$ and the limit essentially saturates. 
This has the unexpected consequence that, 
in the presence of several dark fermion states, 
one can readily get a large suppression of the $B_s$-mixing contribution in the limit where $\sum_{ij} y_s^{i\ast} y_b^i  y_s^{j\ast} y_b^j \ll 1$, which can be easily obtained with, e.g., 
two light states $\chi_1, \chi_2$ and $y_s^{1\ast} y_b^1\approx - y_s^{2\ast} y_b^2$. 

Note that while $C_1$ receives a strong suppression from the addition of approximately equal and opposite-sign 
couplings, there is no equivalent suppression for $C_9^{\mu}$, 
as in the limit of small $x_1, x_2$  
the effective coupling $\tilde{g}$ in Eq.~(\ref{eq:gvbs}) becomes proportional to $\ln (x_1/x_2)$.

\paragraph{\textbf{Limits from $\boldsymbol{B_s}$ decay}} When the axial-vector coupling to the muon, $\gamu$, is present, the vector mediator can induce a contribution to the decay $B_s \to \mu^+ \mu^-$ via the effective operator $\mathcal{O}_6$, cf.~Eq.~\eqref{eq:O6op}. (Equivalently, via the effective $b-s-V$ coupling.) The decay amplitude is expressed in terms of ``effective'' coefficients $C_{10}$ and $C_P$, whose contributions to $B_s \to \mu^+ \mu^-$ are well known~\cite{Chatrchyan:2013bka,CMS:2014xfa,Aaij:2017vad,Aaboud:2018mst}.

By adopting the same convention for the scalar operator as in ref.~\cite{Arnan:2019uhr} we obtain a result similar to~\cite{Sala:2017ihs}:
\begin{eqnarray}
    C_{10} (\mBs^2) &=& \frac{4 \pi \mathcal{N}}{\alpha_{EM}}\frac{\mBs^2 ~\gtilde\, \gamu }{\mBs^2-\mv^2 + i \mv \Gamma_V}\,,\label{eq:c10}\\
    C_P (\mBs^2)  &=& - \frac{ 2 \mmu (M_b + M_s)}{\mv^2}  C_{10} (\mBs^2)\,.\label{eq:cp}
\end{eqnarray}

The typical bounds on $C_P$ are significantly more stringent than 
those on $C_{10}$.
They are likely to have a strong impact on our results so that we 
include them directly in the full numerical scan present in the next section.

\paragraph{\textbf{Limits from $\boldsymbol{B \to K^{(*)}}$ transitions}}

Depending on the details of the UV completion there can exist 
additional $B \to K^{(*)}$ decays providing strong constraints on our model. 
The two main limits are (1) invisible $B \to K$ transition measured by BaBar~\cite{Lees:2013kla,delAmoSanchez:2010bk} and Belle~\cite{Lutz:2013ftz,Grygier:2017tzo},
$\mathcal{BR}(B \to K \nu \nu) \lesssim 1.5 \times 10^{-5}$;  
and (2) resonant search $B \to K^{\ast} V$, 
$V \to \mu \mu$ measured at LHCb~\cite{Aaij:2015tna}, 
which constrains the branching ratio $\mathcal{BR}(B \to K^{\ast}V,\,V \to \mu \mu)\lesssim 2 \times 10^{-9}$.

Given the presence of the heavy scalar doublet $\Phi$ and one dark fermion $\chi_1$, one can construct the tree-level decay process $B\to K\chi_1\chi_1$ based on the $b$ quark 3-body decay $b \to \chi_1 \Phi^{\ast} \to s \chi_1 \chi_1$. In the limit where $m_{\chi_1}, M_K \ll M_B$, we have the simple expression:
\begin{align}
\Gamma^{\textrm{tree}}_{K \chi_1 \chi_1} \approx \frac{f_+^2 |y_s^1|^2 |y_b^1|^2}{1536\, \pi^3} \frac{M_B^5}{m_{\Phi}^4}\,,
\end{align}
where $f_+^2 \approx 0.3$ is the average value of the form factor over the range of integration of the differential decay rate. 
This typically leads to the constraint $|y_s^{1\ast} y_b^1| \lesssim 10^{-2} (m_{\Phi} /\rm{ TeV})^2$ on the Yukawa 
couplings of any new fermion with mass $2m_{\chi_1}<M_B- M_K$. 
As Eq.~(\ref{eq:gvbs}) shows, the upper bound on the Yukawa couplings strongly limits the available range of $g_{Vbs}$, even when the gauge coupling $g_D$ 
is large. This means that to fit the flavour observables one has to resort to a large $\gvmu$ value, which, 
as we shall see in Sec.~\ref{sec:results}, is severely constrained by $Z$-lineshape bounds (in addition to requiring 
a fine tuning of the axial-vector contribution to avoid exceeding $\delta(g-2)_{\mu}$). 
To avoid these problems the minimal particle content will 
have to include at least one dark fermion with mass $m_{\chi_1}>(M_B- M_K)/2$.  

On the other hand, there is a case to be made for the presence of additional light states with mass below the $(M_B- M_K)/2$ threshold and
Yukawa couplings to $\Phi$ and the $b$ and $s$ quarks that are small enough to avoid the tree-level $B\to K+\textrm{inv.}$ bound.
\begin{itemize}
    \item \textit{Dark matter}
    For a fermion $\chi_1$ of mass above the $(M_B- M_K)/2$ threshold the 
    direct $s$-wave annihilation channel, $\chi_1 \chi_1 \to V\to\mu^+ \mu^-$, is strongly constrained by CMB bounds~\cite{Slatyer:2015jla}. Introducing one additional lighter state $\chi_0$ with the same quantum numbers provides instead a viable candidate for \textit{forbidden} dark matter (we come back to this point in more detail at the end of Sec.~\ref{sec:results})
    \item The presence of additional light states directly affects the total width of the gauge boson $V$, potentially opening up additional parameter space for a solution to the flavour anomalies. 
\end{itemize}

In cases where at least one light state $\chi_0$ appears besides $\chi_1$, 
two qualitatively different regimes of applicability should be considered for
the invisible $B \to K$ transition.
For fermions not very light and $\mv < 2 M_{\mu}< M_B - M_K < 2 m_{\chi_{i=0,1}}$,
\textit{on-shell} decay $B\to K V$ occurs, in which $V$ escapes undetected. 
It is typically suppressed in the low $\mv$ limit due to the momentum dependence of the effective coupling $g_{Vbs}$ 
defined in Eq.~(\ref{eq:gvbs}). We get
\begin{equation}
\label{eq:decayKV}
\Gamma_{KV} = f_+^2(\mv) ~\tilde{g}^2 \frac{\mv^2 M_B^3}{64 \pi} \lambda(1,x_K,x_V)^{3/2}\,, 
\end{equation}
where we have assumed $\mv, M_K \ll M_B$, $f_+(0) \simeq 0.3$ is a form factor (the full expression, 
used in the numerical analysis of Sec.~\ref{sec:results}, 
can be found in ref.~\cite{Bailey:2015dka}), 
and $\lambda(x,y,z)=(x-y-z)^2-4xy$ is the standard K\"all\'en (triangle) function with 
$x_K = M_K^2/M_B^2$, $x_V = \mv^2/M_B^2$.

On-shell decays $B\to K\chi_0 \chi_0$, occurring when $2 m_{\chi_0} <\mv < M_B - M_K$, 
require Eq.~(\ref{eq:decayKV}) to be multiplied by the branching ratio $\mathcal{BR}(V\to\chi_0\chi_0)$.
An additional channel is opened via the
exchange of a virtual $V$ and it can dominate the invisible decay width when $\mv$ is small and  the dark coupling $g_D$ is large.
The full width reads
\begin{equation}
\label{eq:decayINV}
\Gamma_{K \chi_0 \chi_0} = \int_{4 \mmu^2}^{(M_B-M_K)^2} \!\!\!\!\!\!\!\! ds ~\frac{\sqrt{s}}{\pi} \frac{\Gamma_{V \to \chi_0 \chi_0} (s)   \Gamma_{KV} (s)}{\mv^2 \Gamma_V^2 + (\mv^2-s)^2}\,,
\end{equation}
where $ \Gamma_{KV} (s)$ and $\Gamma_{V \to \chi_0 \chi_0}$ are obtained by replacing $\mv$ by $\sqrt{s}$ in Eq.~\eqref{eq:decayKV} 
and in the corresponding $V$ decay width to fermions, 
and $\Gamma_V$ is the total width of $V$, cf. Eq.~(\ref{eq:gammav}). 
Note that with more than one dark fermion in the spectrum one ought 
to sum over all individual off-shell contributions.

Altogether, the combination of both real and virtual contribution to the $B\to K\chi_0\chi_0$ decay 
implies a complex kinematic shape in terms of missing energy, 
which may differ significantly from the SM-like $B\to K\nu\nu$ decay. This has the direct consequence that the experimental results of Belle and BaBar, which are optimised for the neutrino process, should not be directly applied to our scenario. We therefore perform a conservative recasting of these analyses, described in detail in Appendix~\ref{appen}. \bigskip

Finally, in the mass regime $\mv > 2 \mmu$, 
the light vector state can directly decay into a muon pair and 
this opens up the resonant channel $B\to K^{*}V$, $V \to \mu \mu$.
This is especially important  
if one restricts the analysis to the case with only one dark fermion $\chi_1$ with $2 m_{\chi_1}>M_B-M_K$. The typical decay width into $K^*$ is given by 
\begin{equation}
\label{eq:decayK}
\Gamma_{K^*V} =  \frac{\tilde{g}^2\mv^2 M_B^5}{64 \pi \mks^2 (1-x_{K^*})^2} \sqrt{\lambda(1,x_{K^*},x_V)} \mathcal{F}_1(x_{K^*},x_V),
\end{equation}
where an explicit expression for $\mathcal{F}_1(x_{K^*},x_V) $ can be found in Appendix~B of ref.~\cite{Altmannshofer:2017bsz}. 
Note that in the limit where $\mv \ll M_B$, $\mathcal{F}_1(x,y)$ simplifies to $\mathcal{F}_1(x,y) \simeq 0.1 \sqrt{x} ( y + 1.2 x) $, so that $\Gamma_{K^*V}$ is typically suppressed compared to  $\Gamma_{KV}$. Furthermore, the branching ratio to muons is inversely proportional to the total width $\Gamma_V$ and can thus be strongly suppressed if the coupling  of $V$ to extra dark fermions $\chi_i$ is large. 

Note that the limits from LHCb~\cite{Aaij:2015tna} on this process focused on a narrow, or even long-lived, new resonance, with limits based on invariant mass bins of a few MeV. This hypothesis is especially problematic when the mediator is around the GeV range, since a large dark gauge coupling implies a very large width for $V$. We then simply model the resonance via a Breit-Wigner distribution and compare bin-by-bin with the limit of ref.~\cite{Aaij:2015tna}, retaining the strongest bin as the main limit.\footnote{While it is clear that a complete recasting of the LHCb analysis~\cite{Aaij:2015tna} for a large-width NP signal would impact the $\sim\textrm{GeV}$ mass range of $V$, we do not expect significant modifications to the overall picture discussed in Sec.~\ref{sec:results} since: (1)~limits on $\mathcal{BR}(B \to K+\rm{inv.})$ \textit{already} forbid scan points compatible with the flavour anomalies in this region when the invisible width is large; (2)~this region has a strong background 
from the charm-quark resonances~\cite{Sala:2017ihs}; 
and (3)~we already include the main flavour observable in the numerical scan via the differential branching ratio of $B \to K \mu \mu$.}

 \begin{figure}[t]
	\centering
		\includegraphics[width=0.75\textwidth]{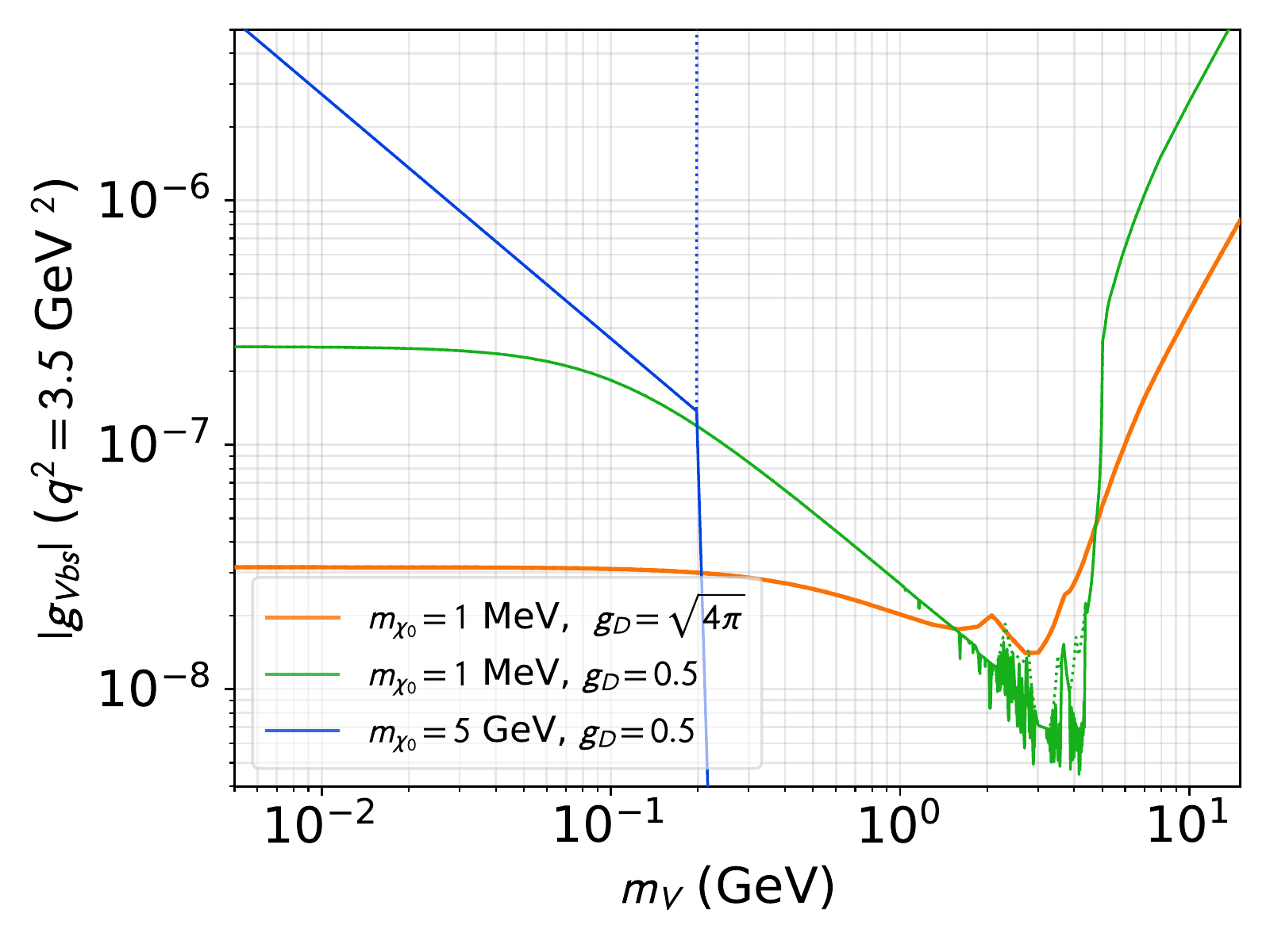}
	\caption{Upper bound on $|g_{Vbs} (3.5\,\textrm{GeV}^2)|$ as a function of $\mv$ for various choices of the input parameters. 
We set $Q_{\Phi}=1$. The specific impact of the different constraints is described in the main text.}
			\label{fig:limitysyb}
\end{figure}

The overall impact of the limits from $B \to K^{(*)}$ transitions on the size of the effective coupling $g_{Vbs}$ 
is summarised in Fig.~\ref{fig:limitysyb} as a function of the vector mass $m_V$, for three representative choices of the pair 
($ m_{\chi_0}$, $g_D$). 
For $m_{\chi_0}=5\,\textrm{GeV}$, 
the decay $B\to K \chi_0 \chi_0$ is kinematically forbidden, 
so that invisible $B$ decays can only proceed via the on-shell process $B\to K V$. 
This results in a weak bound in the small $m_V$ regime (blue line). Note that when $\mv > 2 \mmu$, the constraints from resonant searches using $B \to K^* \mu \mu$ strongly exclude this setup since $V\to\mu\mu$ is in this case the only accessible decay channel.

In the presence of dark fermions with a small mass, $2 m_{\chi_0}< M_B-M_K$, 
the decay channel $V\to \chi_0 \chi_0$ opens up and the relative strength of different bounds 
depends on the size of $g_D$. In the case of large $g_D$ (orange line), if $2 m_{\chi_0} < \mv < M_B-M_K$ both the on-shell and off-shell invisible decays contribute to the width. 
For light vector mediator the off-shell decay takes over due to the $q^2$ dependence of the coupling and the bound on $g_{Vbs}$ saturates. If, on the other hand, $g_D$ is smaller (green line) and $\mv > 2 \mmu$, the constraints from resonant searches using $B \to K^* \mu \mu$ typically overcome the invisible decay limits, suppressing $g_{Vbs}$ by an order-one factor.

Finally, note that 
for large $\mv$ the limit arises from invisible searches from the off-shell decay $B \to K \chi_0 \chi_0$ when it is kinematically allowed.

\subsection{Precision physics constraints\label{sec:prec}}

\paragraph{\textbf{Muon anomalous magnetic moment}}

The couplings of $V$ to muons can be constrained by the measurement of the anomalous magnetic moment of the muon. A contribution to $\gmtwo$ is in this case given by~\cite{Jegerlehner:2009ry,Queiroz:2014zfa}
\begin{equation}
\label{eq:gm2}
\delta(g-2)_\mu =  \frac{1}{8 \pi^2} \frac{\mmu^2}{\mv^2} \mathcal{F}\left(\frac{\mmu}{\mv}\right)\,,
\end{equation}
where 
\begin{equation}\label{eq:gm2loop}
\mathcal{F}(x) =\int_0^1 dz\, \frac{\left(g^V_\mu\right)^2 2 z^2 \left(1-z\right)
+\left(g_\mu^A\right)^2 \left[2z \left(1-z\right) \left(z-4\right) - 4x^2 z ^3\right]}{x^2 z + (1-z)(1-x^2z)}\,.
\end{equation}
\smallskip

As was mentioned in Sec.~\ref{sec:oneloop}, given the limited range achievable in penguin constructions 
for $g_{Vbs}$, when only vector-like couplings to the muons are present it becomes difficult to find a $\gvmu$ value large enough to
allow for a reasonable 
agreement with the flavour anomalies and at the same time not too large a deviation from the measured value of $(g-2)_{\mu}$. 
A certain level of cancellation with the contribution from the axial-vector coupling must
take place in most situations~\cite{Sala:2017ihs}. For a GeV-scale vector mediator, this occurs for $\gamu \approx -0.44\, \gvmu$. Note, however, that including the axial-vector contribution triggers the strong bounds from $B_s \to \mu \mu$ discussed above.

\paragraph{\textbf{$\boldsymbol{Z}$ physics and intensity frontier limits}}

The coupling of the $Z$ boson to the muon is modified at the one-loop level~\cite{Altmannshofer:2014cfa,ALEPH:2005ab} within UV constructions of the type as in Footnote~\ref{foot:uv}.
However, due to the smallness of the $Z$-boson coupling to charged leptons in the SM, the limit is typically subdominant with respect to the $\gmtwo$ bound.

A powerful method for discerning light resonances through precision measurements of Drell-Yan dimuon production was proposed in ref.~\cite{Bishara:2017pje}.
For $m_V=1-5\,\textrm{GeV}$ 
an upper bound can be derived
\begin{align}
\sqrt{(g^V_\mu)^2+(g^A_\mu)^2}\lesssim 5.6\times 10^{-2}\left(1+0.13\,\frac{m_V}{\textrm{GeV}}\right)\,.
\end{align}

Finally, the Belle-II Collaboration recently provided a bound 
on the final state radiation process $e^+ e^- \to \mu^+ \mu V$, $V \to \textrm{invisible}$, based on $0.28\,\textrm{fb}^{-1}$ of data from the 2018 run~\cite{Adachi:2019otg}, which applies directly to our model. 
While the current limit can hardly compete with the Drell-Yan bound, the 2019 run has stored $\sim 10\,\textrm{fb}^{-1}$ and moreover a few $\textrm{ab}^{-1}$ 
should be obtained in 2020, so that future data will become rapidly relevant.

Notice that more generically, for a $V$ mass above $\sim 10\,\textrm{GeV}$, the phenomenology of the light fermions in intensity frontier experiments can be obtained by integrating it out and considering the fermion portal four-fermion operators~\cite{Darme:2020ral}. Similarly, the limit from the tree-level $b \to \chi \Phi^* \to s \chi \chi$ decay can also be obtained  by integrating out the heavy scalar $\Phi$ and using the existing bounds on the fermion portal operator $\bar{b} \gamma^\mu s \bar{\chi} \gamma_\mu \chi$.
While we cover directly the relevant limits in this section, the latter approach could be particularly fruitful to study and constrain the possible couplings 
between new light fermions and the other SM generations.

\paragraph{\textbf{Neutrino trident production}}

If the gauge boson features a coupling $g_{\nu}$ to muon neutrinos (cf. Sec.~\ref{sec:penguin}), 
one expects a strong enhancement in the neutrino trident production from scattering on atomic nuclei, $N$:
$N \nu \rightarrow \nu N \mu^+ \mu^-$~\cite{Altmannshofer:2014pba}. 

The cross section for this process has been measured by the CCFR~\cite{Mishra:1991bv} and CHARM-II~\cite{Geiregat:1990gz} 
collaborations to be in agreement with the SM prediction.
In the range $m_V> 1\,\textrm{GeV}$ it results in the generic bound
\begin{equation}
g_{\nu}\lesssim 0.002 \left(\frac{m_V}{\textrm{GeV}}\right)\,,
\end{equation}
which roughly saturates for smaller mass to about $g_{\nu}\lesssim 0.001$.

When the bound appears in the plots of Sec.~\ref{sec:results} it is obtained under the assumption $g_{\nu}\equiv g_{\mu}^V$ (we repeat that whether or not the bound is relevant depends on the UV completion).

\paragraph{\textbf{Kinetic mixing}}

In presence of states charged both under the U(1)$_Y$ and U(1)$_D$ symmetry groups, kinetic mixing $\epsilon$ between the 
photon and the vector $V$ will be generated at the loop level. 
The corresponding 1-loop contributions from fermions and scalars are given by
\begin{equation}
\label{eq:kinmix}
\epsilon\approx\frac{g_Dg_Y\cos\theta_W}{12\pi^2}\left(\sum_{f\in\textrm{ferm.}}N_{3_f}Y_f Q_f + \frac{1}{8}\sum_{s\in\textrm{scal.}}N_{3_s}Y_s Q_s\right),
\end{equation}
where $Y_{f,s}$ is the fields' hypercharge, $Q_{f,s}$ is the dark charge, and
the coefficients $N_{3_{f,s}}$ indicate the dimension of the SU(3)$_c$ representation. 

The fields that contribute to the kinetic mixing are $\Phi$, $\mu_L$, and $\mu_R$ which, 
when $g_{\mu}^{V,A}\ll g_D$, results in 
\begin{align}
    \epsilon\approx 8 \cdot 10^{-4} \times \left( \frac{g_D}{3}\right)\,.
\end{align}

Such a kinetic mixing is at the limit of exclusion given the current intensity frontier searches (see, e.g., ref.~\cite{Beacham:2019nyx} 
for a recent review), especially when invisible decay channels are not available for the vector mediator. 
However, the precise value of the kinetic mixing is strongly dependent on the UV physics, and additional VL 
fields can modify the prediction of Eq.~\eqref{eq:kinmix} although not by many orders of magnitude.

\paragraph{\textbf{LHC constraints on $t\to c\mu\mu$}}

We work in this paper under the assumption that the only nonzero Yukawa couplings of the down-like type are $y_s^i$, $y_b^i$. However, as 
Eq.~(\ref{eq:L_B}) shows, the corresponding Yukawa couplings of the up-like type, $y_c^i$, $y_t^i$, do not receive CKM~suppression. They can 
thus generate non-negligible contributions to processes involving $t\to c\mu\mu$ transitions.  

Effective operator analyses of LHC bounds from rare top decays~\cite{Durieux:2014xla,Chala:2018agk}, derived originally for 
the very-high mass regime, $m_V\sim\mathcal{O}(\textrm{TeV})$, impose
a fairly weak bound on the coupling product when $m_V\ll M_t$\,:
\begin{equation}
    |\tilde{g}\, \gvmu|\lesssim 10^{-6}\,\textrm{GeV}^{-2}\,.
\end{equation}

A rough comparison with Eq.~(\ref{him_est}) shows that this is not likely 
to be constraining for our scenarios. However, given that  
Eq.~(\ref{eq:Wilson}) presents a nontrivial $q^2$ dependence, for the light mass range investigated in this paper 
one should rather perform a detailed recast of the experimental searches. 
This task exceeds the purpose of the present paper in view of the fact that, as we will show in Sec.~\ref{sec:results}, 
the flavour and intensity frontier experiments discussed above provide already a set of powerful and often inescapable 
constraints on the dark sector. 

\section{Fitting procedure and results\label{sec:results}}

\paragraph{\textbf{Fitting procedure}} 
We perform a multidimensional fit of the following free parameters: 
$m_V$, $g_\mu^V$, $r_{AV}\equiv g_\mu^A/g_\mu^V$, $\gamma_V^D$, $r=g_D Q_\Phi y_s^{1\ast} y_b^1$ (with only one light dark fermion $\chi_1$ contributing with Yukawa couplings $y_s^1$ and $y_b^1$). $r$ is employed here 
as a proxy for the effective coupling $\tilde{g}$ once 
the mass parameters are fixed. We choose $m_{\chi_1} = 2.5\,\textrm{GeV}$ and
$\ms = 1\,\textrm{TeV}$; under these assumptions 
$r$ relates to $\tilde{g}$ as $|\tilde{g}| = |2.2 \times 10^{-8}\,\textrm{GeV}^{-2}\,r|$.
Since the fitted flavour observables depend on $\tilde{g}$ rather than the couplings composing it, our results can be extended straightforwardly to the case with more light fermions.

\begin{table}[!t]
\centering
\renewcommand{\arraystretch}{1.5}
\begin{tabular}{|c|c|}
\hline
\textbf{Parameter} & \textbf{Prior range} \\[1mm]
\hline
$|r|$ & $[0.075, 25]$\\
$g_\mu^V$ & $[0.001, 0.2]$ \\
$\gamma_V^D$ & $[0.001,0.4]$ \\
\hline
$r_{AV}$ & $[-1, 0]$ \\
$m_V (\textrm{GeV})$ & $[0.6, 2]$ \textrm{and} $[2, 15]$ \\
\hline
\end{tabular}
\caption{Prior ranges for the free parameters of the model. For the first three 
parameters the prior distribution 
is given in logarithmic scale, while for the others it is given in linear scale.}
 \label{Tab:Params}
\end{table}

The prior ranges of the fitted parameters are presented in Table~\ref{Tab:Params}.
Separate fits are performed depending on whether $m_V$ lies above or below the relevant bins for the $B$ anomalies. 
In the text we refer to these sets as the \textit{high-mass} fit and the \textit{low-mass} fit. 
This is required by the fact that in order to obtain a negative $C_9^\mu$ the product $r\cdot \gvmu$ should assume a different sign in each of these two regions (we will come back to this point later on, when discussing our numerical results). We observe that $m_V = 2\,\textrm{GeV}$ gives an approximate threshold separating the two regimes.

We fit the free parameters of the model
to the available experimental data reporting anomalies 
in $B$-meson decays, namely the LFUV ratios $R_K$, $R_{K^*}$, the angular observables in the $B \to K^*\mu^+\mu^-$ decay and the branching ratio of $B_s\to \mu\mu$.\footnote{Due 
to the explicit $q^2$ dependence of $C_{9,10}^{\mu}$ it is not possible to match the Wilson coefficients to the model-independent bounds 
obtained by the global fits in the literature. Instead, $C_{9,10}^{\mu}$ have to be fitted directly to the experimental data. On the other hand, if $m_V$ is in the MeV range, it is possible to directly match the Wilson coefficients to the model-independent bounds 
obtained by the global fits.} In the high-mass regime we include in the 
likelihood function the experimental $2 \sigma$  
upper bound on the anomalous magnetic moment of the muon,
$\delta(g-2)_{\mu}\lesssim 4\times 10^{-9}$. We do not incorporate this bound in the likelihood function of the low-mass fit to avoid driving the scan too forcefully towards parameter space regions that are in tension with the remaining constraints.

To carry out the fit, we employ the \texttt{HEPfit} package~\cite{deBlas:2019okz}, performing a Markov Chain Monte Carlo (MCMC) analysis by means of the Bayesian Analysis Toolkit (BAT)~\cite{Caldwell:2008fw}.
A set of $\mathcal{O}(500K)$ points is generated
for the two scenarios described in Table~\ref{Tab:Params}, and for each case the subset of points reproducing the $B$~anomalies, $\mathcal{BR}(B_s\to \mu\mu)$, and the upper $(g-2)_{\mu}$ bound (in the high-mass case only) at the $2 \sigma$ level, is stored. 
These initial subsets of points are then subjected to additional constraints 
coming from $B_s$~mixing,  
Drell-Yan production and, when applicable, $B \to K+\rm{inv.}$ searches.

\paragraph{Results for high-mass} 
We start with addressing solutions in the high-mass range, $m_V=2-15\,\textrm{GeV}$.\smallskip

\begin{figure}[t]
	\centering
		\includegraphics[width=0.75\textwidth]{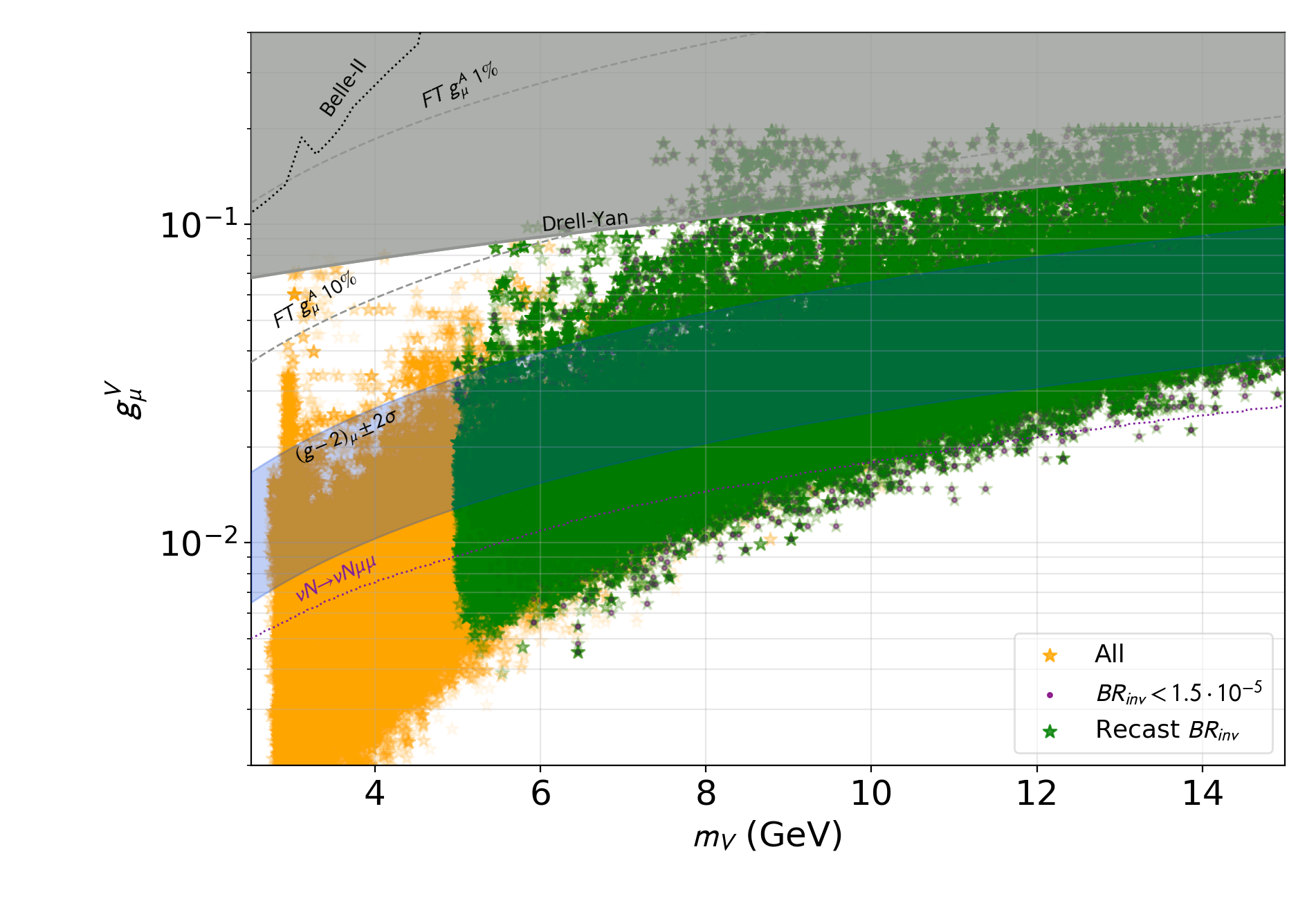}
	\caption{In yellow model points in which the $b\to s$~anomalies, $\mathcal{BR}(B_s\to \mu\mu)$, and $(g-2)_\mu$ are 
	fitted at the $2\sigma$ level. Superimposed points are allowed after the limits from $B \to K+\textrm{inv.}$ are applied, 
	as reported by CLEO~\cite{Ammar:2001gi} (purple) or using our recast procedure (green), cf.~Appendix~\ref{appen}. 
	Grey shaded region is excluded by precision measurements of Drell-Yan at the LHC~\cite{Bishara:2017pje}. 
	Purple dotted line indicates the exclusion bound from the neutrino trident production~\cite{Altmannshofer:2014pba}. Dotted black line indicates the upper bound from Belle~II~\cite{Adachi:2019otg}.
	In light blue we mark the region in which the $\gmtwo$ constraint is satisfied at $2\sigma$ with $\gamu = 0$.}
			\label{fig:ScanHighMV}
\end{figure}

\textit{Two dark fermions in the theory: $m_{\chi_0}\leq m_{\chi_1}$} We describe here a case in which the gauge boson $V$
features a non-negligible invisible width, which could stem from the presence in the spectrum of an additional light fermion $\chi_0$ besides $\chi_1$ (see discussion in Sec.~\ref{sec:flav}).
The results of the scan in the ($m_V$, $\gvmu$) plane are presented in Fig.~\ref{fig:ScanHighMV}.
The yellow points are obtained in the scanning 
procedure described above and correspond to the models in which the 
$B$~anomalies, $\mathcal{BR}(B_s\to \mu\mu)$, and $\delta(g-2)_\mu$ are fitted at the $2\sigma$ level. 
The green points are those that remain allowed after the 
limits from $B \to K+\textrm{inv.}$ transitions are applied, following the recasting procedure outlined in Appendix~\ref{appen}. 

The on-shell process $B \to K V$,  $V \to \chi_0 \chi_0$ typically proceeds unsuppressed for a GeV-scale mediator, so that all the yellow points below the $\mB-\mK$ threshold are excluded. On the other hand, the limits on $\gtilde$ are dramatically weakened above the $\mB-\mK$ threshold (see Fig.~\ref{fig:limitysyb}), 
so that in this regime one can easily fit simultaneously $\gmtwo$ and the flavor anomalies.
Incidentally, we find that solutions to the flavour anomalies with $\mv \approx 2.5\,\textrm{GeV}$, corresponding 
to the case described in ref.~\cite{Sala:2017ihs}, are excluded. The reason is that, unlike ref.~\cite{Sala:2017ihs}, 
in our framework $g_{Vbs}$ is $q^2$-dependent
and induces an $(m_V/\textrm{GeV})^2$ enhancement to the size of Eq.~(\ref{eq:decayKV}). 

An additional bound on the parameter space of the model is derived from 
the $Z$ lineshape from Drell-Yan at the LHC, which strongly affects the maximal allowed value of $g_\mu^V$~\cite{Bishara:2017pje}. 
The corresponding exclusion region is depicted in Fig.~\ref{fig:ScanHighMV} in dark grey. The shading is obtained under 
the assumption $\gamu=-0.44\,\gvmu$, a relation that induces destructive interference in the calculation of $(g-2)_{\mu}$
(cf. Eqs.~(\ref{eq:gm2}) and (\ref{eq:gm2loop})). 
It is well known~\cite{Sala:2017ihs} that the above relation between the vector and axial-vector coupling requires 
some level of fine tuning. 
The grey dashed lines in Fig.~\ref{fig:ScanHighMV} trace the value of $\gvmu$ that corresponds to the indicated level of fine tuning in $\gamu\approx -0.44\,\gvmu$ required to avoid exceeding the $2\sigma$ upper bound from the measurement of $\delta(g-2)_{\mu}$. 
Note, however, that the tuning of the vector and axial-vector muon couplings is a priori not needed in the high-mass regime, 
as confirmed by the large number of green points within the blue shaded band, 
corresponding to the region satisfying the $\gmtwo$ constraint at $2\sigma$ with $\gamu = 0$. 
This is an attractive feature of our model, in which we can obtain relative low values of $g^V_\mu$ 
and subsequently avoid the Drell-Yan limit. 

In Fig.~\ref{fig:ScanHighMV} the limits from the neutrino trident production derived in ref.~\cite{Altmannshofer:2014cfa} for $\gamu = 0$ are shown as a dashed purple line. The constraint applies only if the mediator couples directly to neutrinos, cf. discussion in Sec.~\ref{sec:penguin}. 
It should be stressed that, even when the neutrino trident bound applies, 
solutions that escape the experimental limit exist, with $m_V=5-14\,\textrm{GeV}$ and $r$ at the upper end of the scanned range. 

Finally, it is instructive to compare the results obtained within our UV-complete setup to those derived in simplified $Z'$ models 
with an effective coupling to the $b-s$ current, $g_{bs}$. For example, a solution to the $b-s$~anomalies (without the $\gmtwo$ constraint) was found in ref.~\cite{Alok:2017sui}, with $m_{Z'}=10\,\textrm{GeV}$, $\gvmu=-\gamu\approx 10^{-2}$, and $g_{bs}\approx 5\times 10^{-6}$. Figure~\ref{fig:ScanHighMV} shows that we obtain solutions characterised by similar mass and muon coupling in our setup, corresponding, again, to $r$ values at the upper end of the scanned range.\smallskip

\textit{One dark fermion $\chi_1$ in the theory} 
To conclude the discussion of Fig.~\ref{fig:ScanHighMV}, we point out that the picture does not receive substantial 
modifications if the light fermion $\chi_0$ is not introduced in the theory, and the only NP fermion sits at $m_{\chi_1}=2.5\,\textrm{GeV}$. In this case, the $B\to K+\textrm{inv.}$ bound does not apply. 
However, all points with mass $m_V<M_B-M_K$ become subject to the strong resonant $B\to K^{\ast}\mu\mu$ limit, 
which cuts drastically the parameter space and induces solutions not dissimilar to the area
delimited in green in Fig.~\ref{fig:ScanHighMV}.\smallskip

\textit{Constraints from $B_s$ mixing} The parameter space shown in Fig.~\ref{fig:ScanHighMV} has not been subject to constraints from $B_s$~mixing, which predominantly proceeds via box diagrams involving dark fermions and heavy coloured scalars, and which in principle could put a strong limit on the NP Yukawa couplings. 
On the other hand, the LFUV and angular observables in the fit depend only indirectly on $y_b$ and $y_s$, 
via the effective coupling $\gtilde$. 
In the presence of several light fermions, a possible way of suppressing $B_s$~mixing in the limit $m_{\chi_i}/m_{\Phi}\ll 1$ 
is obtained when $\sum_{ij} y_s^{i\ast} y_b^i  y_s^{j\ast} y_b^j \ll 1$, 
even if one of the $\chi_i$ is relatively heavier than the others, as shown in Sec.~\ref{sec:flav}.

 \begin{figure}[t]
	\centering
		\includegraphics[width=0.75\textwidth]{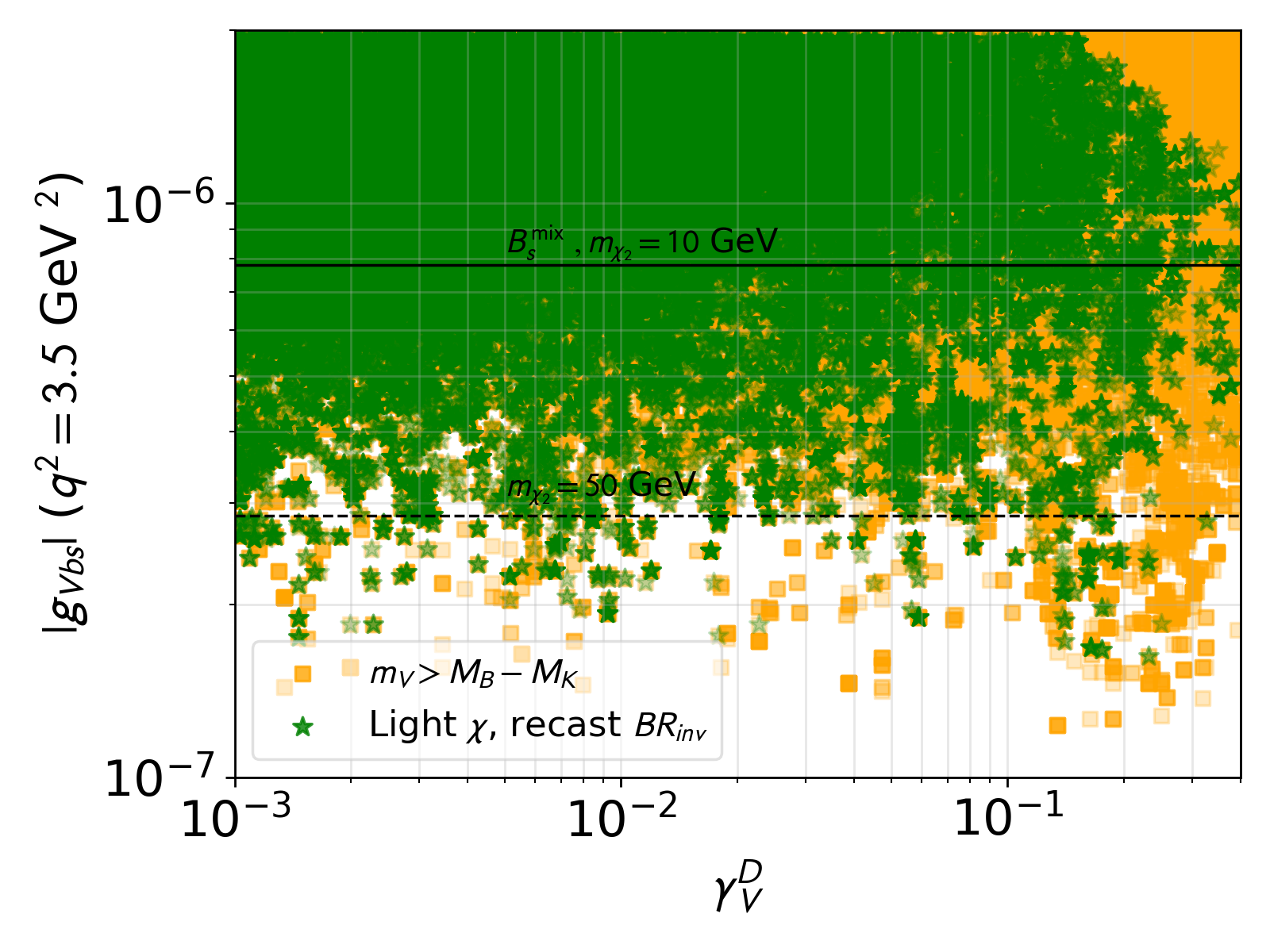}
	\caption{  Results of the scan in the high-mass region, $\mv=2-15\,\textrm{GeV}$,in the plane of the effective coupling $|g_{Vbs}|$ at $q^2=3.5$ GeV$^2$ versus $\gamma_V^D$. All points in the plot satisfy the flavour constraints at $2\sigma$. Yellow squares have $\mv > M_B - M_K$ by default, while green stars lie on a broader mass range but must survive the $B \to K+\textrm{inv.}$ limit. The latter is obtained from our recasting procedure assuming $V$ has a $\sim 100\%$ decay rate into a light dark fermion with $m_{\chi_0} \ll (M_B - M_K)/2$. Horizontal grey lines represent the upper bound from $B_s$~mixing on the effective coupling, under the assumption $y_s^{1\ast} y_b^1 +  y_s^{2\ast} y_b^2 =0$ for two different mass values of the compensating dark fermion: $m_{\chi_2} = 10\,\textrm{GeV}$ (solid) and $m_{\chi_2}= 50\,\textrm{GeV}$ (dashed).}
			\label{fig:rgD}
\end{figure}

We illustrate in Fig.~\ref{fig:rgD} the dependence of the $B_s$-mixing bound, which can limit the size of  $g_{Vbs}$, on the mass
of one additional light fermion introduced in the theory as a means to cancelling the $B_s-B_s$ box diagram. 
We start by ploting $g_{Vbs}$ as a function of the $V$ width, or rather the parameter $\gamma_V^D$ defined in Eq~\eqref{eq:gammavD}.
Green points correspond 
to the solutions also marked in green in  
Fig.~\ref{fig:ScanHighMV}, which provide a satisfying fit to the flavour observables and escape
$B \to K+\textrm{inv.}$ limits when this channel is open. Yellow points show the corresponding case with only $\chi_1$ in the spectrum, which are subject predominantly to the bound from 
$B\to K^{\ast}\mu \mu$. 
The parameter space is partially tilted towards the small width for the green points, since the channel $B\to K+\textrm{inv.}$ is open and constrain the larger invisible widths.

The presence or not of the light state $\chi_0$ has little influence on the overall $B_s$-mixing constraints due to its small Yukawa couplings. On the other hand, the box-diagram induced contribution from $\chi_1$ would limit $g_{Vbs}$ to be around $10^{-8}$, indicating the need for an additional contribution to $B_s$-mixing in the UV of the theory. As we have discussed previously in Sec.~\ref{sec:flav}, a particularly simple way out invokes the presence of an extra, more massive, dark fermion $\chi_2$. The solid line shows the upper bound on $g_{Vbs}$ when $m_{\chi_2}=10\,\textrm{GeV}$, $g_D=\sqrt{4\pi}$,
and we impose $\delta_Y=y_s^{1\ast} y_b^1 +  y_s^{2\ast} y_b^2 =0$. The corresponding limit when
$m_{\chi_2}=50\,\textrm{GeV}$ is shown with a dashed line. Reducing the fine tuning to $\delta_Y=0.05$ lowers the lines by a factor of 5.

 \begin{figure}
	\centering
		\includegraphics[width=0.75\textwidth]{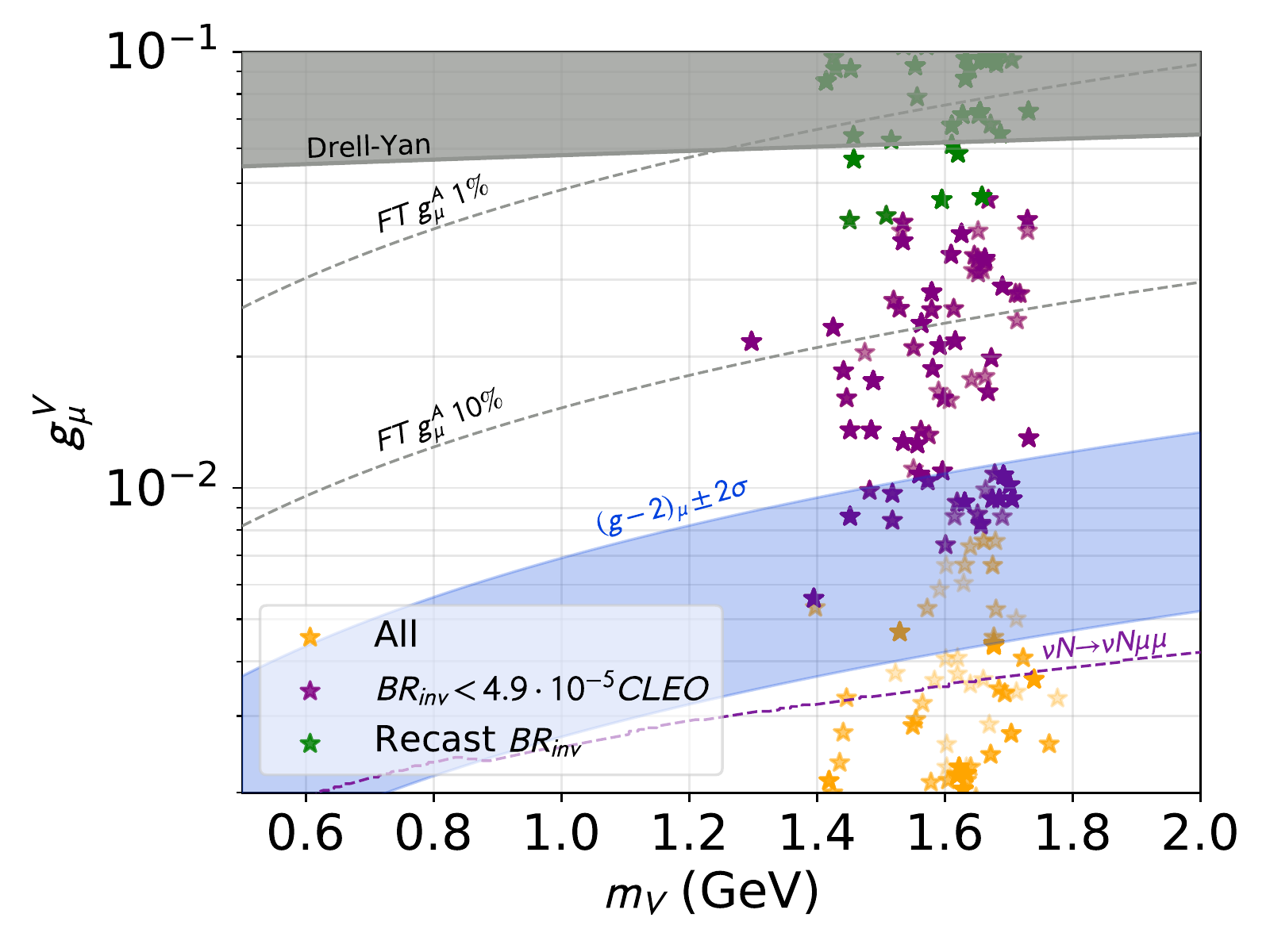}
	\caption{Results of the scan in the low $\mv$ region. The colour code is the same as in Fig.~\ref{fig:ScanHighMV}.}
			\label{fig:ScanLowMV}
\end{figure}

\paragraph{Results for low-mass} We show in Fig.~\ref{fig:ScanLowMV} the results of the scan for the low-mass range, 
$m_V=0.6-2\,\textrm{GeV}$, in the ($m_V$, $\gvmu$) plane. Recall
that to obtain this region of the parameter space one has to switch the sign in the product $r\cdot \gvmu$ with respect to the high-mass region. The procedure allows one to fit $R_{K^{(\ast)}}$ 
correctly by means of destructive interference with the SM value of $C_9^{\mu}$ \textit{below} the experimental bin. 
The colour code in Fig.~\ref{fig:ScanLowMV} is the same as in Fig.~\ref{fig:ScanHighMV}. Note that in this region the mediator \textit{must have} a sizeable invisible width to avoid stringent constraints from a visible dimuon resonance in the $B \to K^*$ spectrum, as discussed in Sec.~\ref{sec:flav}. The presence of at least one light fermion $\chi_0$ besides $\chi_1$ is therefore given for granted.

A few takeaways emerge from the scan in the low-mass region. The first is that there are no 
solutions with mass
$m_V\lesssim 1.4\,\textrm{GeV}$. In fact, they are cut out by the $\mathcal{BR}(B_s\to \mu\mu)$ constraint, 
which is directly implemented in the likelihood function. Besides the mass cut, the surviving points are all 
characterised by fairly large values of $R_{K^{(\ast)}}$, which both lie $2-3\sigma$ away from the central value measured at LHCb 
and closer to their SM expectation. For the same reason, the plot appears much sparser than in the high-mass case: 
very few model points can be found within $2\sigma$ of the measured values of LFUV observables and $\mathcal{BR}(B_s\to \mu\mu)$ simultaneously. 
We thus identify a mild tension in this part of 
the parameter space.  

Note also that the models surviving the bound from $B\to K+\textrm{inv.}$ searches (green points) 
require a large coupling $\gvmu$ to the muon, as
$\tilde{g}$ is directly constrained by the invisible search.
This means a large level of fine tuning in the corresponding $\gamu$ value necessary to cancel 
$\delta(g-2)_{\mu}$. Overall, it is clear that the low-mass region is under siege from a combination 
of complementary bounds but at present it is not entirely excluded.

Let us finish this section by mentioning the case of a very light mediator: a new gauge boson with mass in the MeV range while dark fermions lie in the GeV scale. An interesting property of this regime is that, in the limit where $ 2 \mfi > \mB - \mK$, the mediator $V$ is essentially long-lived since is does not have any available tree-level  decay channel. The invisible $B$ decay is then driven exclusively by the $B \to K V$ process, 
which is strongly suppressed at low $\mv$. Furthermore, the dependence of the Wilson coefficients on $q^2/(q^2 - \mv^2)$ 
converges to a constant when $m_V\ll q$ and it closely resembles the standard electromagnetic penguin contribution to the flavour anomalies.  However, given that the limit from $B_s \to \mu \mu$ decay forbids such a light vector mediator to have a significant axial-vector coupling to the muon, cf. Eqs.~(\ref{eq:c10}), (\ref{eq:cp}), one cannot avoid the upper bound 
on the vector coupling arising from $\delta(g-2)_\mu$\,: $\gvmu \lesssim 7 \times 10^{-4}$.
Concretely, in order to obtain $C_9^\mu \approx -0.7$ a coupling $\gtilde \gtrsim 10^{-6}\,\textrm{GeV}^{-2}$ 
is required, which can only by achieved while satisfying $B \to K+\textrm{inv.}$ limits if $\mv \lesssim 5\,\textrm{MeV}$, cf.~Fig.~\ref{fig:limitysyb}.
A vector mediator this light is already excluded by the standard searches for long-lived dark photon. We conclude that no solution with $V$ in the MeV range is available in penguin-generated scenarios.
 
\paragraph{\textbf{Dark matter}}
As an interesting aside, the lightest dark fermion can provide a good example of \textit{forbidden dark matter} candidate 
when its mass is below the muon mass. The dominant annihilation channel for such a dark matter candidate would be $\chi_0 \bar{\chi}_0 \to V^* \to \mu \bar{\mu}$. Such a process leads to a typical relic density of 
\begin{align}
    \Omega h^2 \approx  0.05 \left( \frac{100 \textrm{ MeV}}{m_{\chi_0}} \right)^2 \left( \frac{\mv} {8 \textrm{ GeV}}\right)^4 \left( \frac{0.05}{\gvmu} \right)^2 \left( \frac{1}{\alpha_D} \right) e^{-2 \Delta\cdot x_f} \ ,
\end{align}
where $x_f \approx 20$ for the relevant masses, $g_D = \sqrt{4 \pi}$ and $\Delta = 1 - m_{\chi_0}/\mmu$.

When $m_{\chi_0}$ drops below the muon mass threshold the relic density is exponentially enhanced since the annihilation process can only occur due to the thermal velocity of the dark matter particle in the early universe~\cite{Griest:1990kh,DAgnolo:2015ujb}. 
This ensures that the thermal target is matched for one coupling-dependent mass below $\mmu$, typically around $\sim \mmu/2$. Furthermore, all other annihilation processes are exponentially suppressed when the universe temperature decreases, ensuring that the CMB limits on late-time annihilating sub-GeV dark matter are automatically escaped. 

\section{Conclusions}\label{sec:concl}

We have presented in this work a solution for the $b \to s$ flavour anomalies based on the presence of a split dark sector with a light vector mediator as well as new light Dirac fermions which may constitute all or part of the dark matter. The interaction with the $b$ and $s$ quarks is generated at the loop level via the addition of a coloured scalar particle, resembling a supersymmetric squark. We analysed numerically and analytically the resulting low-energy effective theory, which in particular possesses a $q^2$-dependent interaction of the vector mediator with $b$ and $s$ quarks. Varying the mass of the vector mediator from the MeV scale to the tens of GeV, we find two scenarios satisfying all experimental constraints while providing a good fit to the anomalies. In particular, the region with a GeV-scale mediator above the $B-K$ mass threshold appears particularly promising, requiring little to no tuning in the low-energy effective parameters, 
and to the best of our knowledge it has not been considered previously.

Since our model is partially embedded in a UV~completion, we have additionally pointed out several constraints 
that can challenge its viability. 
We have highlighted the constraints from the $B\to K+\textrm{inv.}$ decay rate and 
$B_s$~mixing and, in the latter case, provided an example of a mechanism to escape it. 
We did not make any assumption in this paper on the nature of the dark Dirac-fermion interactions with the neutrinos. Indeed, since the former are complete SM singlets, it would be very interesting to investigate whether or not they could behave as right-handed neutrinos (for instance via the coupling to a dark charged new Higgs doublet), and in that case investigate
their relationship with the strong neutrino trident limits.

While the experimental constraints on models addressing the flavour anomalies with light mediators are already quite stringent, we have 
identified several observables that can easily exclude these scenarios entirely or provide smoking-gun proof of their detection. Chief among those are the limits from $B \to K$ and $B \to K^*$ transitions. While the former play a critical role via the $B \to K+\textrm{inv.}$ bounds, experimental searches are typically optimised for the SM process $B \to K \nu \nu$. 
Including an analysis based on a light, and potentially broad, invisible resonance $B \to K V$ could likely strengthen significantly the existing limits, especially when the mediator is light. Similarly, the latest search for $B \to K^* \mu \mu$ 
has focused on a very narrow resonance, and should be properly recast for the case of a large and invisible width. 
Finally, it is important to note that limits on a light dark photon are due to improve in the next few years, and will further constrain the case of a mediator at and below the GeV scale.

\bigskip
\noindent \textbf{Acknowledgments}
\medskip

\noindent
LD thanks S. Robertson, M. Heck, M. Williams and G. De Pietro for interesting discussions. We thank Pere Arnan for pointing out a factor of 2 mistake in the normalisation of Eq.~\eqref{eq:gtilde}. LD is supported by the INFN “Iniziativa Specifica” Theoretical Astroparticle Physics (TAsP-LNF). MF is supported by the MINECO grant FPA2016-76005-C2-1-P and by Maria de Maetzu program grant MDM-2014-0367 of ICCUB and 2017 SGR 929. KK is supported in part by the National Science  Centre (Poland)  under  the  research  Grant No. 2017/26/E/ST2/00470. EMS is supported in part by  the  National  Science  Centre  (Poland)  under  the research Grant No. 2017/26/D/ST2/00490. 

\newpage
\appendix

\section{Appendix: Invisible decay limits}\label{appen}

We present in this appendix a more detailed treatment of the recasting procedure performed to extract conservative $B \to K +\textrm{inv.}$ 
limits on our model.

While the CLEO Collaboration searched explicitly for an on-shell light particle mediating the $B \to K$ decay~\cite{Ammar:2001gi}, 
their $2\sigma$ limit $\mathcal{BR}(B \to K+\textrm{inv.}) < 4.9 \times 10^{-5}$ is relatively weak compared 
to the ones from $B$~factories. In the following we will base our limit both on the BaBar result~\cite{Lees:2013kla} -- which provides a differential branching ratio limit in bins of $q^2$ -- as well as on an older analysis from the same collaboration~\cite{delAmoSanchez:2010bk}, which also had some differential limits, albeit on a much larger range for~$q^2$. Note that the current bounds from the Belle Collaboration~\cite{Lutz:2013ftz,Grygier:2017tzo} are typically of the same order as for BaBar. 
They are however strongly optimised for the SM-like $B \to K \nu \nu$ signal
and only present their bounds in the total integrated branching ratio. We therefore concentrate on the two BaBar analyses.

First, using the BaBar hadronic-tagging analysis~\cite{Lees:2013kla}, we calculate the branching ratio $\mathcal{BR}(B \to K \chi_1 \chi_1)$ in $s_B = q^2 / M_B^2$ bins, where for low $\mv$ most of our NP signal is concentrated in the lowest bin, $s_B < 0.1$. 
We then compare it with the $2\sigma$ limits from Fig.~6a of ref.~\cite{Lees:2013kla}. While this approach leads to a strong bound 
when the real process $B \to K V$, $V\to \chi_0 \chi_0$ dominates, these limits can be significantly weakened when the virtual process dominates, since the branching ratio accounts for a broader spread in $s_B$ bins.

We therefore also include partially integrated limits from the BaBar semileptonic-tagging analysis~\cite{delAmoSanchez:2010bk}, which combined the world-leading limit on $B^+ \to K^+ \nu \nu$ with detailed information about the signal efficiencies as function of the momentum of the $K^+$ (and hence on the missing energy). We select the $q^2$ ranges $[3.4^2,\,4^2]\,\textrm{GeV}^2$ 
and $[0,\,2.4^2]\,\textrm{GeV}^2$ (corresponding to $p_K$ in the range $[1,\,1.5]\,\textrm{GeV}$ 
and~$\gtrsim 2\,\textrm{GeV}$, respectively), where the Boosted Decision Tree~(BDT) efficiencies presented in Fig.~3 of ref.~\cite{delAmoSanchez:2010bk} are larger than $\sim 0.3$. This ensures that the signal efficiencies for our 
NP kinematics are of the same order of magnitude or higher than the ones for the SM signal. 
We then compare both regions with the low-$q^2$ and high-$q^2$ 95\%~C.L. limits,  
$\mathcal{BR}(B \to K+\textrm{inv.}) < 1.1 \times 10^{-5}$ 
and $\mathcal{BR}(B \to K+\textrm{inv.}) < 4.6 \times 10^{-5}$, respectively.

\bigskip
\bibliographystyle{JHEP}
\bibliography{HSFlavour}

\providecommand{\href}[2]{#2}\begingroup\raggedright\begin{thebibliography}{10}

\bibitem{Aaij:2014ora}
{\bf LHCb} Collaboration, R.~Aaij et~al., {\it {Test of lepton universality
  using $B^{+}\rightarrow K^{+}\ell^{+}\ell^{-}$ decays}},  {\em Phys. Rev.
  Lett.} {\bf 113} (2014) 151601, [\href{http://arxiv.org/abs/1406.6482}{{\tt
  arXiv:1406.6482}}].

\bibitem{Aaij:2017vbb}
{\bf LHCb} Collaboration, R.~Aaij et~al., {\it {Test of lepton universality
  with $B^{0} \rightarrow K^{*0}\ell^{+}\ell^{-}$ decays}},  {\em JHEP} {\bf
  08} (2017) 055, [\href{http://arxiv.org/abs/1705.05802}{{\tt
  arXiv:1705.05802}}].

\bibitem{Aaij:2019wad}
{\bf LHCb} Collaboration, R.~Aaij et~al., {\it {Search for lepton-universality
  violation in $B^+\to K^+\ell^+\ell^-$ decays}},
  \href{http://arxiv.org/abs/1903.09252}{{\tt arXiv:1903.09252}}.

\bibitem{Abdesselam:2019wac}
{\bf Belle} Collaboration, A.~Abdesselam et~al., {\it {Test of lepton flavor
  universality in ${B\to K^\ast\ell^+\ell^-}$ decays at Belle}},
  \href{http://arxiv.org/abs/1904.02440}{{\tt arXiv:1904.02440}}.

\bibitem{Aaij:2015oid}
{\bf LHCb} Collaboration, R.~Aaij et~al., {\it {Angular analysis of the $B^{0}
  \to K^{*0} \mu^{+} \mu^{-}$ decay using 3 fb$^{-1}$ of integrated
  luminosity}},  {\em JHEP} {\bf 02} (2016) 104,
  [\href{http://arxiv.org/abs/1512.04442}{{\tt arXiv:1512.04442}}].

\bibitem{Wehle:2016yoi}
{\bf Belle} Collaboration, S.~Wehle et~al., {\it {Lepton-Flavor-Dependent
  Angular Analysis of $B\to K^\ast \ell^+\ell^-$}},  {\em Phys. Rev. Lett.}
  {\bf 118} (2017), no.~11 111801, [\href{http://arxiv.org/abs/1612.05014}{{\tt
  arXiv:1612.05014}}].

\bibitem{Sirunyan:2017dhj}
{\bf CMS} Collaboration, A.~M. Sirunyan et~al., {\it {Measurement of angular
  parameters from the decay $\mathrm{B}^0 \to \mathrm{K}^{*0} \mu^+ \mu^-$ in
  proton-proton collisions at $\sqrt{s} = $ 8 TeV}},  {\em Phys. Lett.} {\bf
  B781} (2018) 517--541, [\href{http://arxiv.org/abs/1710.02846}{{\tt
  arXiv:1710.02846}}].

\bibitem{Aaboud:2018krd}
{\bf ATLAS} Collaboration, M.~Aaboud et~al., {\it {Angular analysis of $B^0_d
  \rightarrow K^{*}\mu^+\mu^-$ decays in $pp$ collisions at $\sqrt{s}= 8$ TeV
  with the ATLAS detector}},  {\em JHEP} {\bf 10} (2018) 047,
  [\href{http://arxiv.org/abs/1805.04000}{{\tt arXiv:1805.04000}}].

\bibitem{Aaij:2015esa}
{\bf LHCb} Collaboration, R.~Aaij et~al., {\it {Angular analysis and
  differential branching fraction of the decay $B^0_s\to\phi\mu^+\mu^-$}},
  {\em JHEP} {\bf 09} (2015) 179, [\href{http://arxiv.org/abs/1506.08777}{{\tt
  arXiv:1506.08777}}].

\bibitem{Aaij:2014pli}
{\bf LHCb} Collaboration, R.~Aaij et~al., {\it {Differential branching
  fractions and isospin asymmetries of $B \to K^{(*)} \mu^+ \mu^-$ decays}},
  {\em JHEP} {\bf 06} (2014) 133, [\href{http://arxiv.org/abs/1403.8044}{{\tt
  arXiv:1403.8044}}].

\bibitem{Aaij:2016flj}
{\bf LHCb} Collaboration, R.~Aaij et~al., {\it {Measurements of the S-wave
  fraction in $B^{0}\rightarrow K^{+}\pi^{-}\mu^{+}\mu^{-}$ decays and the
  $B^{0}\rightarrow K^{\ast}(892)^{0}\mu^{+}\mu^{-}$ differential branching
  fraction}},  {\em JHEP} {\bf 11} (2016) 047,
  [\href{http://arxiv.org/abs/1606.04731}{{\tt arXiv:1606.04731}}]. [Erratum:
  JHEP04,142(2017)].

\bibitem{Altmannshofer:2014rta}
W.~Altmannshofer and D.~M. Straub, {\it {New physics in $b\rightarrow s$
  transitions after LHC run 1}},  {\em Eur. Phys. J.} {\bf C75} (2015), no.~8
  382, [\href{http://arxiv.org/abs/1411.3161}{{\tt arXiv:1411.3161}}].

\bibitem{Altmannshofer:2017fio}
W.~Altmannshofer, C.~Niehoff, P.~Stangl, and D.~M. Straub, {\it {Status of the
  $B\rightarrow K^*\mu ^+\mu ^-$ anomaly after Moriond 2017}},  {\em Eur. Phys.
  J.} {\bf C77} (2017), no.~6 377, [\href{http://arxiv.org/abs/1703.09189}{{\tt
  arXiv:1703.09189}}].

\bibitem{Capdevila:2017bsm}
B.~Capdevila, A.~Crivellin, S.~Descotes-Genon, J.~Matias, and J.~Virto, {\it
  {Patterns of New Physics in $b\to s\ell^+\ell^-$ transitions in the light of
  recent data}},  {\em JHEP} {\bf 01} (2018) 093,
  [\href{http://arxiv.org/abs/1704.05340}{{\tt arXiv:1704.05340}}].

\bibitem{Altmannshofer:2017yso}
W.~Altmannshofer, P.~Stangl, and D.~M. Straub, {\it {Interpreting Hints for
  Lepton Flavor Universality Violation}},  {\em Phys. Rev.} {\bf D96} (2017),
  no.~5 055008, [\href{http://arxiv.org/abs/1704.05435}{{\tt
  arXiv:1704.05435}}].

\bibitem{DAmico:2017mtc}
G.~D'Amico, M.~Nardecchia, P.~Panci, F.~Sannino, A.~Strumia, R.~Torre, and
  A.~Urbano, {\it {Flavour anomalies after the $R_{K^*}$ measurement}},  {\em
  JHEP} {\bf 09} (2017) 010, [\href{http://arxiv.org/abs/1704.05438}{{\tt
  arXiv:1704.05438}}].

\bibitem{Ciuchini:2017mik}
M.~Ciuchini, A.~M. Coutinho, M.~Fedele, E.~Franco, A.~Paul, L.~Silvestrini, and
  M.~Valli, {\it {On Flavourful Easter eggs for New Physics hunger and Lepton
  Flavour Universality violation}},  {\em Eur. Phys. J.} {\bf C77} (2017),
  no.~10 688, [\href{http://arxiv.org/abs/1704.05447}{{\tt arXiv:1704.05447}}].

\bibitem{Alok:2017sui}
A.~K. Alok, B.~Bhattacharya, A.~Datta, D.~Kumar, J.~Kumar, and D.~London, {\it
  {New Physics in $b \to s \mu^+ \mu^-$ after the Measurement of $R_{K^*}$}},
  {\em Phys. Rev.} {\bf D96} (2017), no.~9 095009,
  [\href{http://arxiv.org/abs/1704.07397}{{\tt arXiv:1704.07397}}].

\bibitem{Hurth:2014vma}
T.~Hurth, F.~Mahmoudi, and S.~Neshatpour, {\it {Global fits to $b \to
  s\ell\ell$ data and signs for lepton non-universality}},  {\em JHEP} {\bf 12}
  (2014) 053, [\href{http://arxiv.org/abs/1410.4545}{{\tt arXiv:1410.4545}}].

\bibitem{Hurth:2016fbr}
T.~Hurth, F.~Mahmoudi, and S.~Neshatpour, {\it {On the anomalies in the latest
  LHCb data}},  {\em Nucl. Phys.} {\bf B909} (2016) 737--777,
  [\href{http://arxiv.org/abs/1603.00865}{{\tt arXiv:1603.00865}}].

\bibitem{Chobanova:2017ghn}
V.~G. Chobanova, T.~Hurth, F.~Mahmoudi, D.~Martinez~Santos, and S.~Neshatpour,
  {\it {Large hadronic power corrections or new physics in the rare decay $B\to
  K^{\ast}\mu^{+}\mu^{-}$?}},  {\em JHEP} {\bf 07} (2017) 025,
  [\href{http://arxiv.org/abs/1702.02234}{{\tt arXiv:1702.02234}}].

\bibitem{Hurth:2017hxg}
T.~Hurth, F.~Mahmoudi, D.~Martinez~Santos, and S.~Neshatpour, {\it {Lepton
  nonuniversality in exclusive $b{\rightarrow}s{\ell}{\ell}$ decays}},  {\em
  Phys. Rev.} {\bf D96} (2017), no.~9 095034,
  [\href{http://arxiv.org/abs/1705.06274}{{\tt arXiv:1705.06274}}].

\bibitem{Arbey:2018ics}
A.~Arbey, T.~Hurth, F.~Mahmoudi, and S.~Neshatpour, {\it {Hadronic and New
  Physics Contributions to $b \to s$ Transitions}},  {\em Phys. Rev.} {\bf D98}
  (2018), no.~9 095027, [\href{http://arxiv.org/abs/1806.02791}{{\tt
  arXiv:1806.02791}}].

\bibitem{Alguero:2019ptt}
M.~Algueró, B.~Capdevila, A.~Crivellin, S.~Descotes-Genon, P.~Masjuan,
  J.~Matias, and J.~Virto, {\it {Emerging patterns of New Physics with and
  without Lepton Flavour Universal contributions}},  {\em Eur. Phys. J.} {\bf
  C79} (2019), no.~8 714, [\href{http://arxiv.org/abs/1903.09578}{{\tt
  arXiv:1903.09578}}].

\bibitem{Alok:2019ufo}
A.~K. Alok, A.~Dighe, S.~Gangal, and D.~Kumar, {\it {Continuing search for new
  physics in $b \to s \mu \mu$ decays: two operators at a time}},  {\em JHEP}
  {\bf 06} (2019) 089, [\href{http://arxiv.org/abs/1903.09617}{{\tt
  arXiv:1903.09617}}].

\bibitem{Ciuchini:2019usw}
M.~Ciuchini, A.~M. Coutinho, M.~Fedele, E.~Franco, A.~Paul, L.~Silvestrini, and
  M.~Valli, {\it {New Physics in $b \to s \ell^+ \ell^-$ confronts new data on
  Lepton Universality}},  {\em Eur. Phys. J.} {\bf C79} (2019), no.~8 719,
  [\href{http://arxiv.org/abs/1903.09632}{{\tt arXiv:1903.09632}}].

\bibitem{Datta:2019zca}
A.~Datta, J.~Kumar, and D.~London, {\it {The $B$ anomalies and new physics in
  $b \to s e^+ e^-$}},  {\em Phys. Lett.} {\bf B797} (2019) 134858,
  [\href{http://arxiv.org/abs/1903.10086}{{\tt arXiv:1903.10086}}].

\bibitem{Aebischer:2019mlg}
J.~Aebischer, W.~Altmannshofer, D.~Guadagnoli, M.~Reboud, P.~Stangl, and D.~M.
  Straub, {\it {B-decay discrepancies after Moriond 2019}},
  \href{http://arxiv.org/abs/1903.10434}{{\tt arXiv:1903.10434}}.

\bibitem{Kowalska:2019ley}
K.~Kowalska, D.~Kumar, and E.~M. Sessolo, {\it {Implications for new physics in
  $b\rightarrow s \mu \mu $ transitions after recent measurements by Belle and
  LHCb}},  {\em Eur. Phys. J.} {\bf C79} (2019), no.~10 840,
  [\href{http://arxiv.org/abs/1903.10932}{{\tt arXiv:1903.10932}}].

\bibitem{Arbey:2019duh}
A.~Arbey, T.~Hurth, F.~Mahmoudi, D.~M. Santos, and S.~Neshatpour, {\it {Update
  on the $b\to s$ anomalies}},  {\em Phys. Rev.} {\bf D100} (2019), no.~1
  015045, [\href{http://arxiv.org/abs/1904.08399}{{\tt arXiv:1904.08399}}].

\bibitem{Bhattacharya:2019dot}
S.~Bhattacharya, A.~Biswas, S.~Nandi, and S.~K. Patra, {\it {Exhaustive Model
  Selection in $b \to s \ell \ell$ Decays: Pitting Cross-Validation against
  AIC$_c$}},  \href{http://arxiv.org/abs/1908.04835}{{\tt arXiv:1908.04835}}.

\bibitem{Datta:2017pfz}
A.~Datta, J.~Liao, and D.~Marfatia, {\it {A light $Z^\prime$ for the $R_K$
  puzzle and nonstandard neutrino interactions}},  {\em Phys. Lett.} {\bf B768}
  (2017) 265--269, [\href{http://arxiv.org/abs/1702.01099}{{\tt
  arXiv:1702.01099}}].

\bibitem{Sala:2017ihs}
F.~Sala and D.~M. Straub, {\it {A New Light Particle in B Decays?}},  {\em
  Phys. Lett.} {\bf B774} (2017) 205--209,
  [\href{http://arxiv.org/abs/1704.06188}{{\tt arXiv:1704.06188}}].

\bibitem{Datta:2017ezo}
A.~Datta, J.~Kumar, J.~Liao, and D.~Marfatia, {\it {New light mediators for the
  $R_K$ and $R_{K^*}$ puzzles}},  {\em Phys. Rev.} {\bf D97} (2018), no.~11
  115038, [\href{http://arxiv.org/abs/1705.08423}{{\tt arXiv:1705.08423}}].

\bibitem{Altmannshofer:2017bsz}
W.~Altmannshofer, M.~J. Baker, S.~Gori, R.~Harnik, M.~Pospelov, E.~Stamou, and
  A.~Thamm, {\it {Light resonances and the low-q$^{2}$ bin of $ {R}_{K^{*}}
  $}},  {\em JHEP} {\bf 03} (2018) 188,
  [\href{http://arxiv.org/abs/1711.07494}{{\tt arXiv:1711.07494}}].

\bibitem{Datta:2019bzu}
A.~Datta, J.~L. Feng, S.~Kamali, and J.~Kumar, {\it {Resolving the
  $(g-2)_{\mu}$ and $B$ Anomalies with Leptoquarks and a Dark Higgs Boson}},
  {\em Phys. Rev.} {\bf D101} (2020), no.~3 035010,
  [\href{http://arxiv.org/abs/1908.08625}{{\tt arXiv:1908.08625}}].

\bibitem{Lyon:2014hpa}
J.~Lyon and R.~Zwicky, {\it {Resonances gone topsy turvy - the charm of QCD or
  new physics in $b \to s \ell^+ \ell^-$?}},
  \href{http://arxiv.org/abs/1406.0566}{{\tt arXiv:1406.0566}}.

\bibitem{Aaij:2015dea}
{\bf LHCb} Collaboration, R.~Aaij et~al., {\it {Angular analysis of the $B^{0}
  \to K^{*0} e^{+} e^{-}$ decay in the low-q$^{2}$ region}},  {\em JHEP} {\bf
  04} (2015) 064, [\href{http://arxiv.org/abs/1501.03038}{{\tt
  arXiv:1501.03038}}].

\bibitem{Gripaios:2015gra}
B.~Gripaios, M.~Nardecchia, and S.~A. Renner, {\it {Linear flavour violation
  and anomalies in B physics}},  {\em JHEP} {\bf 06} (2016) 083,
  [\href{http://arxiv.org/abs/1509.05020}{{\tt arXiv:1509.05020}}].

\bibitem{Arnan:2016cpy}
P.~Arnan, L.~Hofer, F.~Mescia, and A.~Crivellin, {\it {Loop effects of heavy
  new scalars and fermions in $b\to s\mu^+\mu^-$}},  {\em JHEP} {\bf 04} (2017)
  043, [\href{http://arxiv.org/abs/1608.07832}{{\tt arXiv:1608.07832}}].

\bibitem{Cline:2017qqu}
J.~M. Cline and J.~M. Cornell, {\it {$R({K^{(*)}})$ from dark matter
  exchange}},  {\em Phys. Lett.} {\bf B782} (2018) 232--237,
  [\href{http://arxiv.org/abs/1711.10770}{{\tt arXiv:1711.10770}}].

\bibitem{Crivellin:2018yvo}
A.~Crivellin, C.~Greub, D.~M{\"u}ller, and F.~Saturnino, {\it {Importance of
  Loop Effects in Explaining the Accumulated Evidence for New Physics in B
  Decays with a Vector Leptoquark}},  {\em Phys. Rev. Lett.} {\bf 122} (2019),
  no.~1 011805, [\href{http://arxiv.org/abs/1807.02068}{{\tt
  arXiv:1807.02068}}].

\bibitem{Datta:2018xty}
A.~Datta, B.~Dutta, S.~Liao, D.~Marfatia, and L.~E. Strigari, {\it {Neutrino
  scattering and B anomalies from hidden sector portals}},  {\em JHEP} {\bf 01}
  (2019) 091, [\href{http://arxiv.org/abs/1808.02611}{{\tt arXiv:1808.02611}}].

\bibitem{Barman:2018jhz}
B.~Barman, D.~Borah, L.~Mukherjee, and S.~Nandi, {\it {Correlating the
  anomalous results in $b \to s$ decays with inert Higgs doublet dark matter
  and muon $(g-2)$}},  {\em Phys. Rev.} {\bf D100} (2019), no.~11 115010,
  [\href{http://arxiv.org/abs/1808.06639}{{\tt arXiv:1808.06639}}].

\bibitem{Marzo:2019ldg}
C.~Marzo, L.~Marzola, and M.~Raidal, {\it {Common explanation to the
  $R_{K^{(*)}}$, $R_{D^{(*)}}$ and $\epsilon^\prime/\epsilon$ anomalies in a
  3HDM+$\nu_R$ and connections to neutrino physics}},  {\em Phys. Rev.} {\bf
  D100} (2019), no.~5 055031, [\href{http://arxiv.org/abs/1901.08290}{{\tt
  arXiv:1901.08290}}].

\bibitem{Arnan:2019uhr}
P.~Arnan, A.~Crivellin, M.~Fedele, and F.~Mescia, {\it {Generic loop effects of
  new scalars and fermions in $b\to s\ell^+\ell^-$ and a vector-like $4^{\rm
  th}$ generation}},  {\em JHEP} {\bf 06} (2019) 118,
  [\href{http://arxiv.org/abs/1904.05890}{{\tt arXiv:1904.05890}}].

\bibitem{Kawamura:2017ecz}
J.~Kawamura, S.~Okawa, and Y.~Omura, {\it {Interplay between the b$\to
  s\ell\ell$ anomalies and dark matter physics}},  {\em Phys. Rev.} {\bf D96}
  (2017), no.~7 075041, [\href{http://arxiv.org/abs/1706.04344}{{\tt
  arXiv:1706.04344}}].

\bibitem{Aaboud:2017wqg}
{\bf ATLAS} Collaboration, M.~Aaboud et~al., {\it {Search for supersymmetry in
  events with $b$-tagged jets and missing transverse momentum in $pp$
  collisions at $\sqrt{s}=13$ TeV with the ATLAS detector}},  {\em JHEP} {\bf
  11} (2017) 195, [\href{http://arxiv.org/abs/1708.09266}{{\tt
  arXiv:1708.09266}}].

\bibitem{Aaboud:2017ayj}
{\bf ATLAS} Collaboration, M.~Aaboud et~al., {\it {Search for a scalar partner
  of the top quark in the jets plus missing transverse momentum final state at
  $\sqrt{s}$=13 TeV with the ATLAS detector}},  {\em JHEP} {\bf 12} (2017) 085,
  [\href{http://arxiv.org/abs/1709.04183}{{\tt arXiv:1709.04183}}].

\bibitem{Sirunyan:2019glc}
{\bf CMS} Collaboration, A.~M. Sirunyan et~al., {\it {Search for direct top
  squark pair production in events with one lepton, jets, and missing
  transverse momentum at 13 TeV with the CMS experiment}},
  \href{http://arxiv.org/abs/1912.08887}{{\tt arXiv:1912.08887}}.

\bibitem{Sirunyan:2019ctn}
{\bf CMS} Collaboration, A.~M. Sirunyan et~al., {\it {Search for supersymmetry
  in proton-proton collisions at 13 TeV in final states with jets and missing
  transverse momentum}},  {\em JHEP} {\bf 10} (2019) 244,
  [\href{http://arxiv.org/abs/1908.04722}{{\tt arXiv:1908.04722}}].

\bibitem{DiLuzio:2017fdq}
L.~Di~Luzio, M.~Kirk, and A.~Lenz, {\it {Updated $B_s$-mixing constraints on
  new physics models for $b\to s\ell^+\ell^-$ anomalies}},  {\em Phys. Rev.}
  {\bf D97} (2018), no.~9 095035, [\href{http://arxiv.org/abs/1712.06572}{{\tt
  arXiv:1712.06572}}].

\bibitem{Kowalska:2017iqv}
K.~Kowalska and E.~M. Sessolo, {\it {Expectations for the muon g-2 in
  simplified models with dark matter}},  {\em JHEP} {\bf 09} (2017) 112,
  [\href{http://arxiv.org/abs/1707.00753}{{\tt arXiv:1707.00753}}].

\bibitem{Sirunyan:2018lul}
{\bf CMS} Collaboration, A.~M. Sirunyan et~al., {\it {Searches for pair
  production of charginos and top squarks in final states with two oppositely
  charged leptons in proton-proton collisions at $\sqrt{s}=$ 13 TeV}},  {\em
  JHEP} {\bf 11} (2018) 079, [\href{http://arxiv.org/abs/1807.07799}{{\tt
  arXiv:1807.07799}}].

\bibitem{Aad:2019vnb}
{\bf ATLAS} Collaboration, G.~Aad et~al., {\it {Search for electroweak
  production of charginos and sleptons decaying into final states with two
  leptons and missing transverse momentum in $\sqrt{s}=13$ TeV $pp$ collisions
  using the ATLAS detector}},  {\em Eur. Phys. J.} {\bf C80} (2020), no.~2 123,
  [\href{http://arxiv.org/abs/1908.08215}{{\tt arXiv:1908.08215}}].

\bibitem{Bennett:2006fi}
{\bf Muon g-2} Collaboration, G.~W. Bennett et~al., {\it {Final Report of the
  Muon E821 Anomalous Magnetic Moment Measurement at BNL}},  {\em Phys. Rev.}
  {\bf D73} (2006) 072003, [\href{http://arxiv.org/abs/hep-ex/0602035}{{\tt
  hep-ex/0602035}}].

\bibitem{Davier:2016iru}
M.~Davier, {\it {Update of the Hadronic Vacuum Polarisation Contribution to the
  muon g-2}},  {\em Nucl. Part. Phys. Proc.} {\bf 287-288} (2017) 70--75,
  [\href{http://arxiv.org/abs/1612.02743}{{\tt arXiv:1612.02743}}].

\bibitem{Jegerlehner:2017lbd}
F.~Jegerlehner, {\it {Muon g-2 theory: The hadronic part}},  {\em EPJ Web
  Conf.} {\bf 166} (2018) 00022, [\href{http://arxiv.org/abs/1705.00263}{{\tt
  arXiv:1705.00263}}].

\bibitem{Altmannshofer:2014pba}
W.~Altmannshofer, S.~Gori, M.~Pospelov, and I.~Yavin, {\it {Neutrino Trident
  Production: A Powerful Probe of New Physics with Neutrino Beams}},  {\em
  Phys. Rev. Lett.} {\bf 113} (2014) 091801,
  [\href{http://arxiv.org/abs/1406.2332}{{\tt arXiv:1406.2332}}].

\bibitem{Mishra:1991bv}
{\bf CCFR} Collaboration, S.~R. Mishra et~al., {\it {Neutrino tridents and W Z
  interference}},  {\em Phys. Rev. Lett.} {\bf 66} (1991) 3117--3120.

\bibitem{Geiregat:1990gz}
{\bf CHARM-II} Collaboration, D.~Geiregat et~al., {\it {First observation of
  neutrino trident production}},  {\em Phys. Lett.} {\bf B245} (1990) 271--275.

\bibitem{Ko:2013zsa}
P.~Ko, Y.~Omura, and C.~Yu, {\it Higgs phenomenology in type-i 2hdm with
  $u(1)\_h$ higgs gauge symmetry},  {\em JHEP} {\bf 01} (2014) 016,
  [\href{http://arxiv.org/abs/1309.7156}{{\tt arXiv:1309.7156}}].

\bibitem{Gracey:2000am}
J.~Gracey, {\it {Three loop MS-bar tensor current anomalous dimension in QCD}},
   {\em Phys. Lett. B} {\bf 488} (2000) 175--181,
  [\href{http://arxiv.org/abs/hep-ph/0007171}{{\tt hep-ph/0007171}}].

\bibitem{Chatrchyan:2013bka}
{\bf CMS} Collaboration, S.~Chatrchyan et~al., {\it Measurement of the $b^0\_s
  \to \mu^+ \mu^-$ branching fraction and search for $b^0 \to \mu^+ \mu^-$ with
  the cms experiment},  {\em Phys.Rev.Lett.} {\bf 111} (2013) 101804,
  [\href{http://arxiv.org/abs/1307.5025}{{\tt arXiv:1307.5025}}].

\bibitem{CMS:2014xfa}
{\bf CMS, LHCb} Collaboration, V.~Khachatryan et~al., {\it Observation of the
  rare $b^0\_s\to\mu^+\mu^-$ decay from the combined analysis of cms and lhcb
  data},  {\em Nature} {\bf 522} (2015) 68--72,
  [\href{http://arxiv.org/abs/1411.4413}{{\tt arXiv:1411.4413}}].

\bibitem{Aaij:2017vad}
{\bf LHCb} Collaboration, R.~Aaij et~al., {\it Measurement of the
  $b^0\_s\to\mu^+\mu^-$ branching fraction and effective lifetime and search
  for $b^0\to\mu^+\mu^-$ decays},  {\em Phys.Rev.Lett.} {\bf 118} (2017),
  no.~19 191801, [\href{http://arxiv.org/abs/1703.05747}{{\tt
  arXiv:1703.05747}}].

\bibitem{Aaboud:2018mst}
{\bf ATLAS} Collaboration, M.~Aaboud et~al., {\it Study of the rare decays of
  $b^0\_s$ and $b^0$ mesons into muon pairs using data collected during 2015
  and 2016 with the atlas detector},  {\em JHEP} {\bf 04} (2019) 098,
  [\href{http://arxiv.org/abs/1812.03017}{{\tt arXiv:1812.03017}}].

\bibitem{Lees:2013kla}
{\bf BaBar} Collaboration, J.~P. Lees et~al., {\it {Search for $B \to K^{(*)}
  \nu \overline \nu$ and invisible quarkonium decays}},  {\em Phys. Rev.} {\bf
  D87} (2013), no.~11 112005, [\href{http://arxiv.org/abs/1303.7465}{{\tt
  arXiv:1303.7465}}].

\bibitem{delAmoSanchez:2010bk}
{\bf BaBar} Collaboration, P.~del Amo~Sanchez et~al., {\it {Search for the Rare
  Decay $B \to K \nu \bar{\nu}$}},  {\em Phys. Rev.} {\bf D82} (2010) 112002,
  [\href{http://arxiv.org/abs/1009.1529}{{\tt arXiv:1009.1529}}].

\bibitem{Lutz:2013ftz}
{\bf Belle} Collaboration, O.~Lutz et~al., {\it {Search for $B \to h^{(*)} \nu
  \bar{\nu}$ with the full Belle $\Upsilon(4S)$ data sample}},  {\em Phys.
  Rev.} {\bf D87} (2013), no.~11 111103,
  [\href{http://arxiv.org/abs/1303.3719}{{\tt arXiv:1303.3719}}].

\bibitem{Grygier:2017tzo}
{\bf Belle} Collaboration, J.~Grygier et~al., {\it {Search for
  $\boldsymbol{B\to h\nu\bar{\nu}}$ decays with semileptonic tagging at
  Belle}},  {\em Phys. Rev.} {\bf D96} (2017), no.~9 091101,
  [\href{http://arxiv.org/abs/1702.03224}{{\tt arXiv:1702.03224}}]. [Addendum:
  Phys. Rev.D97,no.9,099902(2018)].

\bibitem{Aaij:2015tna}
{\bf LHCb} Collaboration, R.~Aaij et~al., {\it {Search for hidden-sector bosons
  in $B^0 \!\to K^{*0}\mu^+\mu^-$ decays}},  {\em Phys. Rev. Lett.} {\bf 115}
  (2015), no.~16 161802, [\href{http://arxiv.org/abs/1508.04094}{{\tt
  arXiv:1508.04094}}].

\bibitem{Slatyer:2015jla}
T.~R. Slatyer, {\it {Indirect dark matter signatures in the cosmic dark ages.
  I. Generalizing the bound on s-wave dark matter annihilation from Planck
  results}},  {\em Phys. Rev.} {\bf D93} (2016), no.~2 023527,
  [\href{http://arxiv.org/abs/1506.03811}{{\tt arXiv:1506.03811}}].

\bibitem{Bailey:2015dka}
J.~A. Bailey et~al., {\it {$B\to Kl^+l^-$ Decay Form Factors from Three-Flavor
  Lattice QCD}},  {\em Phys. Rev.} {\bf D93} (2016), no.~2 025026,
  [\href{http://arxiv.org/abs/1509.06235}{{\tt arXiv:1509.06235}}].

\bibitem{Jegerlehner:2009ry}
F.~Jegerlehner and A.~Nyffeler, {\it {The Muon g-2}},  {\em Phys. Rept.} {\bf
  477} (2009) 1--110, [\href{http://arxiv.org/abs/0902.3360}{{\tt
  arXiv:0902.3360}}].

\bibitem{Queiroz:2014zfa}
F.~S. Queiroz and W.~Shepherd, {\it {New Physics Contributions to the Muon
  Anomalous Magnetic Moment: A Numerical Code}},  {\em Phys. Rev.} {\bf D89}
  (2014), no.~9 095024, [\href{http://arxiv.org/abs/1403.2309}{{\tt
  arXiv:1403.2309}}].

\bibitem{Altmannshofer:2014cfa}
W.~Altmannshofer, S.~Gori, M.~Pospelov, and I.~Yavin, {\it {Quark flavor
  transitions in $L_\mu-L_\tau$ models}},  {\em Phys. Rev.} {\bf D89} (2014)
  095033, [\href{http://arxiv.org/abs/1403.1269}{{\tt arXiv:1403.1269}}].

\bibitem{ALEPH:2005ab}
{\bf ALEPH, DELPHI, L3, OPAL, SLD, LEP Electroweak Working Group, SLD
  Electroweak Group, SLD Heavy Flavour Group} Collaboration, S.~Schael et~al.,
  {\it {Precision electroweak measurements on the $Z$ resonance}},  {\em Phys.
  Rept.} {\bf 427} (2006) 257--454,
  [\href{http://arxiv.org/abs/hep-ex/0509008}{{\tt hep-ex/0509008}}].

\bibitem{Bishara:2017pje}
F.~Bishara, U.~Haisch, and P.~F. Monni, {\it {Regarding light resonance
  interpretations of the B decay anomalies}},  {\em Phys. Rev.} {\bf D96}
  (2017), no.~5 055002, [\href{http://arxiv.org/abs/1705.03465}{{\tt
  arXiv:1705.03465}}].

\bibitem{Adachi:2019otg}
{\bf Belle-II} Collaboration, I.~Adachi et~al., {\it {Search for an Invisibly
  Decaying $Z^{\prime}$ Boson at Belle II in $e^+ e^- \to \mu^+ \mu^- (e^{\pm}
  \mu^{\mp})$ Plus Missing Energy Final States}},
  \href{http://arxiv.org/abs/1912.11276}{{\tt arXiv:1912.11276}}.

\bibitem{Darme:2020ral}
L.~Darmé, S.~A.~R. Ellis, and T.~You, {\it {Light Dark Sectors through the
  Fermion Portal}},  \href{http://arxiv.org/abs/2001.01490}{{\tt
  arXiv:2001.01490}}.

\bibitem{Beacham:2019nyx}
J.~Beacham et~al., {\it {Physics Beyond Colliders at CERN: Beyond the Standard
  Model Working Group Report}},  {\em J. Phys.} {\bf G47} (2020), no.~1 010501,
  [\href{http://arxiv.org/abs/1901.09966}{{\tt arXiv:1901.09966}}].

\bibitem{Durieux:2014xla}
G.~Durieux, F.~Maltoni, and C.~Zhang, {\it {Global approach to top-quark
  flavor-changing interactions}},  {\em Phys. Rev.} {\bf D91} (2015), no.~7
  074017, [\href{http://arxiv.org/abs/1412.7166}{{\tt arXiv:1412.7166}}].

\bibitem{Chala:2018agk}
M.~Chala, J.~Santiago, and M.~Spannowsky, {\it {Constraining four-fermion
  operators using rare top decays}},
  \href{http://arxiv.org/abs/1809.09624}{{\tt arXiv:1809.09624}}.

\bibitem{deBlas:2019okz}
J.~De~Blas et~al., {\it $\texttt{HEPfit}$: a code for the combination of
  indirect and direct constraints on high energy physics models},
  \href{http://arxiv.org/abs/1910.14012}{{\tt arXiv:1910.14012}}.

\bibitem{Caldwell:2008fw}
A.~Caldwell, D.~Kollar, and K.~Kroninger, {\it {BAT: The Bayesian Analysis
  Toolkit}},  {\em Comput. Phys. Commun.} {\bf 180} (2009) 2197--2209,
  [\href{http://arxiv.org/abs/0808.2552}{{\tt arXiv:0808.2552}}].

\bibitem{Ammar:2001gi}
{\bf CLEO} Collaboration, R.~Ammar et~al., {\it {Search for the familon via
  $B^\pm \to \pi^\pm X^0$, $B^\pm \to K^\pm X^0$ , and $B^0 \to K^0_S X^0$
  decays}},  {\em Phys. Rev. Lett.} {\bf 87} (2001) 271801,
  [\href{http://arxiv.org/abs/hep-ex/0106038}{{\tt hep-ex/0106038}}].

\bibitem{Griest:1990kh}
K.~Griest and D.~Seckel, {\it {Three exceptions in the calculation of relic
  abundances}},  {\em Phys. Rev.} {\bf D43} (1991) 3191--3203.

\bibitem{DAgnolo:2015ujb}
R.~T. D'Agnolo and J.~T. Ruderman, {\it {Light Dark Matter from Forbidden
  Channels}},  {\em Phys. Rev. Lett.} {\bf 115} (2015), no.~6 061301,
  [\href{http://arxiv.org/abs/1505.07107}{{\tt arXiv:1505.07107}}].

\end{thebibliography}\endgroup

\end{document}